\RequirePackage{ifpdf}
\ifpdf 
\documentclass[pdftex]{sigma}
\else
\documentclass{sigma}
\fi

\begin{document}

\renewcommand{\textfraction}{0.1}
\renewcommand{\topfraction}{0.9}

\allowdisplaybreaks

\renewcommand{\PaperNumber}{079}

\FirstPageHeading

\ShortArticleName{Clif\/ford Algebras and Possible Kinematics}

\ArticleName{Clif\/ford Algebras and Possible Kinematics}

\Author{Alan S. MCRAE}

\AuthorNameForHeading{A.S.~McRae}

\Address{Department of Mathematics, Washington and Lee University, Lexington, VA  24450-0303, USA }
\Email{\href{mailto:mcraea@wlu.edu}{mcraea@wlu.edu}}

\ArticleDates{Received April 30, 2007, in f\/inal form July 03, 2007; Published online July 19, 2007}

\Abstract{We review Bacry and L\'evy-Leblond's
work on possible kinematics as applied to 2-dimensional spacetimes,
as well as the nine types of 2-dimensional Cayley--Klein geo\-met\-ries,
illustrating how the Cayley--Klein geo\-met\-ries give homogeneous
spacetimes for all but one of the kinematical groups.
We then construct a two-parameter family of Clif\/ford algebras
that give a unif\/ied framework for representing both the Lie algebras
as well as the kinematical groups, showing that these groups are true
rotation groups.  In addition we give conformal models for these spacetimes.}


\Classification{11E88; 15A66; 53A17}

\newcommand{\ka}{{\kappa_1}}
\newcommand{\kb}{{\kappa_2}}
\newcommand{\sia}{{\sigma_2}}
\newcommand{\sib}{{\sigma_3}}
\newcommand{\sic}{{\sigma_1}}
\newcommand{\Ka}{{K_1}}
\newcommand{\Kb}{{K_2}}
\newcommand{\Q}{{\mathbb{Q}}}
\newcommand{\R}{{\mathbb{R}}}
\newcommand{\C}{{\mathbb{C}}}
\newcommand{\CA}{{\C_{\ka}}}
\newcommand{\CB}{{\C_{\kb}}}
\newcommand{\CAB}{{\C_{\ka \kb}}}
\newcommand{\SO}{{SO_{\ka,\kb}(3)}}
\newcommand{\so}{{so_{\ka,\kb}(3)}}
\newcommand{\sO}{{S_{\left[ \ka \right],\kb}^2}}
\newcommand{\sOO}{{S_{\ka, \left[ \kb \right]}^2}}
\newcommand{\Ca}{{C_{\ka}}}
\newcommand{\Cb}{{C_{\kb}}}
\newcommand{\Cab}{{C_{\ka \kb}}}
\newcommand{\Sa}{{S_{\ka}}}
\newcommand{\Sb}{{S_{\kb}}}
\newcommand{\Sab}{{S_{\ka \kb}}}
\newcommand{\Cc}{{C_{\kappa}}}
\newcommand{\Sc}{{S_{\kappa}}}
\newcommand{\Tc}{{T_{\kappa}}}
\newcommand{\Ta}{{T_{\ka}}}
\newcommand{\Tb}{{T_{\kb}}}
\newcommand{\Tab}{{T_{\ka \kb}}}
\newcommand{\wa}{{\mathsf{w}}}
\newcommand{\wb}{{\mathfrak{w}}}
\newcommand{\ns}{{\hat{n} \cdot \vec{\sigma}}}

 \begin{flushright}
 \begin{minipage}{15cm}
 \small \it
\noindent As long as algebra and geometry have been separated,
their progress have been slow and their uses limited; but when
these two sciences have been united, they have lent each mutual
forces, and have marched together towards perfection.\\
\null \hfill Joseph Louis Lagrange (1736--1813)
 \end{minipage}
\end{flushright}

\bigskip

\noindent
The f\/irst part of this paper is a review of Bacry
and L\'evy-Leblond's description of possible kinematics and how
such kinematical structures relate to the Cayley--Klein formalism.
We review some of the work done by Ballesteros, Herranz, Ortega
and Santander on homogeneous spaces, as this work gives a unif\/ied
and detailed description of possible kinematics (save for static kinematics).
The second part builds on this work by analyzing the corresponding kinematical
models from other unif\/ied viewpoints, f\/irst through generalized complex matrix
realizations and then through a two-parameter family of Clif\/ford algebras.
These parameters are the same as those given by Ballesteros et.~al.,
and relate to the speed of light and the universe time radius.

\pdfbookmark[1]{Part~I. A review of kinematics via Cayley-Klein geometries}{part1}
\section*{Part~I. A review of kinematics via Cayley--Klein geometries}


\section{Possible kinematics}

As noted by Inonu and Wigner in their work \cite{IW53}
on contractions of groups and their representations,
classical mechanics is a limiting case of relativistic mechanics,
for both the Galilei group as well as its Lie algebra are limits of the
Poincar\'{e} group and its Lie algebra.  Bacry and L\'{e}vy-Leblond
\cite{BL67} classif\/ied and investigated the nature of
all possible Lie algebras for kinema\-tical groups (these
groups are assumed to be Lie groups as 4-dimensional spacetime
is assumed to be continuous) given the three basic principles that
\begin{itemize}\itemsep=0pt
\item[(i)] space is isotropic and spacetime is homogeneous,
\item[(ii)] parity and time-reversal are automorphisms of the kinematical group, and
\item[(iii)] the one-dimensional subgroups generated by the boosts are non-compact.
\end{itemize}

\begin{table}[th]
 \centering
 \caption{The 11 possible kinematical groups.}
 \vspace{1mm}
  \begin{tabular}{ c | c  }
  \hline
Symbol &  Name  \tsep{1ex} \bsep{1ex}\\
\hline
\tsep{1ex} $dS_1$             & de Sitter group $SO(4,1)$ \\
$dS_2$             & de Sitter group $SO(3,2)$ \\
$P$                   & Poincar\'{e} group \\
$P^{\prime}_1$ & Euclidean group $SO(4)$ \\
$P^{\prime}_2$ & Para-Poincar\'{e} group \\
$C$                    &Carroll group \\
$N_+$               & Expanding Newtonian Universe group \\
$N_-$                & Oscillating Newtonian Universe group \\
$G$                   & Galilei group \\
$G^{\prime}$     & Para-Galilei group \\
$St$                   & Static Universe group \bsep{0.5ex}\\
\hline
 \end{tabular}
 \end{table}

\begin{figure}[th]
\begin{center}
\includegraphics[width=12.5cm]{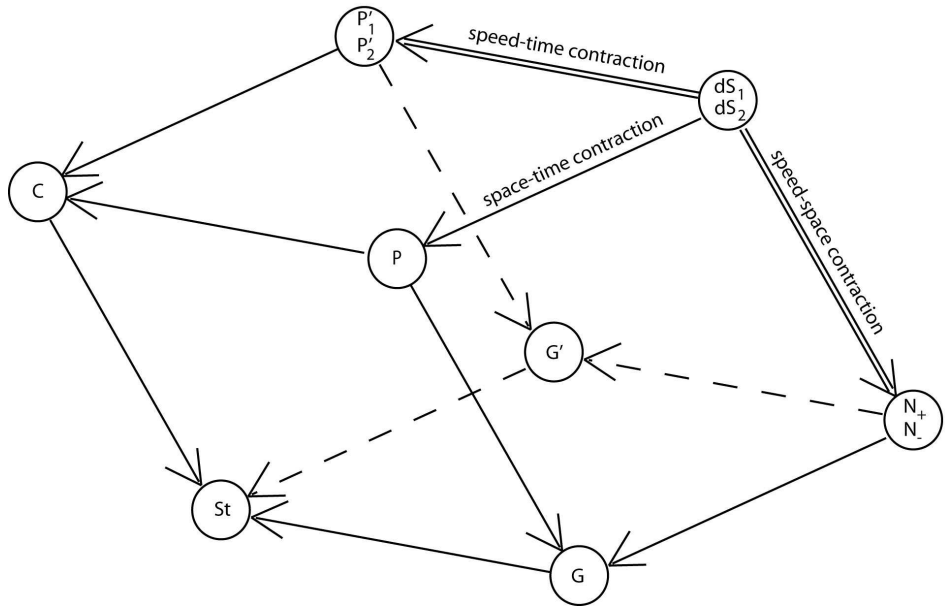}
\end{center}
\caption{The contractions of the kinematical groups.}
\end{figure}

The resulting possible Lie algebras give 11 possible kinematics,
where each of the kinematical groups (see Table~1) is generated by its inertial
transformations as well as its spacetime translations and spatial rotations.
These groups consist of the de Sitter groups and their
rotation-invariant contractions:  the physical nature of
a contracted group is determined by the nature of the contraction
itself, along with the nature of the parent de Sitter group.
Below we will illustrate the nature of these contractions when
we look more closely at the simpler case of a~2-dimensional spacetime.
For Fig.~1, note that a ``upper" face of the cube is transformed under
one type of contraction into the opposite face.

Sanjuan \cite{F84} noted that the methods employed by Bacry
and L\'{e}vy-Leblond could be easily applied to 2-dimensional
spacetimes: as it is the purpose of this paper to investigate
these kinematical Lie algebras and groups through Clif\/ford algebras,
we will begin by explicitly classifying all such possible Lie algebras.
This section then is a detailed and expository account of certain parts of Bacry,
L\'{e}vy-Leblond, and Sanjuan's work.

Let $K$ denote the generator of the inertial transformations, $H$ the generator
of time translations, and $P$ the generator of space translations.
As space is one-dimensional, space is isotropic.   In the following
section we will see how to construct, for each possible kinematical
structure, a~spacetime that is a homogeneous space for its kinematical
group, so that basic principle (i) is satisf\/ied.

Now let $\Pi$ and $\Theta$ denote the respective operations
of parity and time-reversal:  $K$ must be odd under both $\Pi$ and $\Theta$.
Our basic principle (ii) requires that the Lie algebra is acted upon by
the $\mathbb{Z}_2 \otimes \mathbb{Z}_2$ group of involutions generated by
\[ \Pi \, : \, \left( K, H, P \right) \rightarrow \left( -K, H, -P \right) \qquad
\mbox{and} \qquad  \Theta \, : \, \left( K, H, P \right) \rightarrow \left( -K, -H, P \right) .\]

Finally, basic principle (iii) requires that the subgroup
generated by $K$ is noncompact, even though we will allow for
the universe to be closed, or even for closed time-like worldlines to exist.
We do not wish for $e^{0K} = e^{\theta K}$ for some non-zero $\theta$,
for then we would f\/ind it possible for a boost to be no boost at all!

As each Lie bracket $[K, H]$, $[K, P]$, and $[H, P]$ is invariant
under the involutions $\Pi$ and $\Theta$ as well as the involution
\[ \Gamma = \Pi\Theta \, : \, \left( K, H, P \right)
\rightarrow \left( K, -H, -P \right), \]
we must have that $[K, H] = pP$, $[K, P] = hH$,
and $[H, P] = kK$ for some constants $k$, $h$, and $p$.
Note that these Lie brackets are also invariant under the symmetries def\/ined by
\begin{gather*}
 S_P : \{ K \leftrightarrow H, p \leftrightarrow -p, k \leftrightarrow h \}, \qquad
 S_H : \{ K \leftrightarrow P, h \leftrightarrow -h, k \leftrightarrow -p \}, \qquad
\mbox{and}\\
 S_K : \{ H \leftrightarrow P, k \leftrightarrow -k, h \leftrightarrow p \} ,
 \end{gather*}
and that the Jacobi identity is automatically satisf\/ied
for any triple of elements of the Lie algebra.

\begin{table}[t]
 \centering
  \caption{The 21 kinematical Lie algebras, grouped into 11 essentially distinct types of kinematics.}
 \vspace{1mm}

 \begin{tabular}{| r | r | r | r | r | r | r | r | r | r | r | r | r | r | r | r |}
 \cline{1-2} \cline{4-5} \cline{7-8} \cline{10-11} \cline{13-14} \cline{16-16}
 $P$ & $-P$ & & $P$ & $-P$ & & $P$ & $-P$ & & $0$  & $0$    & & $0$ & $0$  & & $0$ \\
 $H$ & $-H$ & & $H$ & $-H$ & & $0$ & $0$   & & $H$ & $-H$  & & $0$ & $0$   & & $0$\\
 $K$ & $-K$ & & $-K$ & $K$ & & $0$ & $0$   & & $0$  & $0$     & & $K$ & $-K$ & & $0$ \\
 \cline{1-2} \cline{4-5} \cline{7-8} \cline{10-11} \cline{13-14} \cline{16-16}
  \multicolumn{16}{c}{}       \\[-2mm]
  \cline{1-2} \cline{4-5} \cline{7-8} \cline{10-11} \cline{13-14}
 $0$ & $0$   & & $0$  & $0$  & & $P$ & $-P$ & & $P$ & $-P$    & & $P$ & $-P$ & \multicolumn{2}{c}{} \\
 $H$ & $-H$ & & $H$ & $-H$ & & $0$ & $0$ & & $0$   & $0$     & & $H$ & $-H$ & \multicolumn{2}{c}{} \\
 $K$ & $-K$ & & $-K$ & $K$ & & $K$ & $-K$ & & $-K$ & $K$    & & $0$ & $0$   & \multicolumn{2}{c}{} \\
 \cline{1-2} \cline{4-5} \cline{7-8} \cline{10-11} \cline{13-14}
 \end{tabular}
 \end{table}

   \begin{table}[t]
 \centering
 \caption{6 non-kinematical Lie algebras.}
 \vspace{1mm}
   \begin{tabular}{| r | r | r | r | r | r | r | r | }  \cline{1-2} \cline{4-5} \cline{7-8}
 $P$ & $-P$ & & $P$ & $-P$ & & $-P$ & $P$ \\
 $-H$ & $H$ & & $-H$ & $H$ & & $H$ & $-H$ \\
 $K$ & $-K$ & & $-K$ & $K$  & & $0$ & $0$ \\ \cline{1-2} \cline{4-5} \cline{7-8}
 \end{tabular}\vspace{-2mm}
 \end{table}

We can normalize the constants $k$, $h$, and $p$ by
a scale change so that $k, h, p \in \{-1, 0, 1 \}$,
taking advantage of the simple form of the Lie brackets
for the basis elements $K$, $H$, and $P$.   There
are then $3^3$ possible Lie algebras, which we
tabulate in Tables~2 and~3 with columns that have the following form:
\begin{table}[htbp]
 \centering
 \begin{tabular}{| r | }  \hline
 $[K, H]$ \\
 $[K, P]$ \\
 $[H, P]$ \\\hline
 \end{tabular}
 \end{table}

  \begin{table}[t]
 \centering
 \caption{Some kinematical groups along with their notation and structure constants.}
\vspace{1mm}
  \begin{tabular}{ c c c c c } \hline
             &  Anti-de Sitter & Oscillating Newtonian Universe & Para-Minkowski      & Minkowski   \\  \cline{2-5}
             &\tsep{0.2ex} $adS$            & $N_-$                          & $M^{\prime}$           & $M$ \\ \hline \hline
\tsep{0.2ex}$[K,H]$ & $P$                & $P$                              &  $0$                        &$P$ \\
$[K,P]$ & $H$                & $0$                              & $H$                         & $H$ \\
$[H,P]$ & $K$                & $K$                             & $K$                          & $0$ \\
\hline
 \end{tabular}
 \end{table}

\begin{table}[t]
 \centering
 \caption{Some kinematical groups along with their notation and structure constants.}
\vspace{1mm}
  \begin{tabular}{ c c c c  } \hline
             &  de Sitter       & Expanding Newtonian Universe & Expanding Minkowski Universe     \\   \cline{2-4}
             & $dS$            & $N_+$      & $M_+$   \\  \hline \hline
\tsep{0.2ex}$[K,H]$ & $P$                & $P$         &  $0$           \\
$[K,P]$ & $H$                & $0$         & $H$               \\
$[H,P]$ & $-K$                & $-K$        & $-K$           \\ \hline
 \end{tabular}
  \end{table}

 \begin{table}[t]
 \centering
\caption{Some kinematical groups along with their notation and structure constants.}
 \vspace{1mm}
 \begin{tabular}{ c c c c c } \hline
             & Galilei            & Carroll    & Static de Sitter Universe   & Static Universe  \\  \cline{2-5}
             & $G$               & $C$         & $SdS$         & $St$ \\ \hline \hline
$[K,H]$ & $P$                & $0$         &  $0$           &$0$ \\
$[K,P]$ & $0$                & $H$         & $0$                & $0$ \\
$[H,P]$ & $0$                & $0$        & $K$          & $0$ \\ \hline
 \end{tabular}
 \end{table}

  \noindent We also pair each Lie
 algebra with its image under the isomorphism given by
 $P \leftrightarrow -P$, $H \leftrightarrow -H$,
 $K \leftrightarrow -K$, and $[ \star, \star \star] \leftrightarrow [ \star \star, \star]$,
 for both Lie algebras then give the same kinematics.  There are then 11 essentially
 distinct kinematics, as illustrated in Table~2.
 Also (as we shall see in the next section) each of the other 6 Lie algebras
 (that are given in Table~3) violate the third basic principle,
 generating a compact group of inertial transformations.

These non-kinematical Lie algebras are the lie algebras for
the motion groups for the elliptic, hyperbolic, and Euclidean planes:
let us denote these respective groups as $El$, $H$, and $Eu$.

We name the kinematical groups (that are generated by
the boosts and translations) in concert with the 4-dimensional case (see Tables~4,~5, and~6).
Each of these kinematical groups is either the de Sitter or the anti-de Sitter group,
or one of their contractions.  We can contract with respect to any subgroup,
giving us three fundamental types of contraction: {\it speed-space, speed-time},
and {\it space-time contractions}, corresponding respectively to contracting
to the subgroups generated by $H$, $P$, and $K$.

{\bfseries \itshape Speed-space contractions.}
We make the substitutions $K \rightarrow \epsilon K$ and
$P \rightarrow \epsilon P$ into the Lie algebra and then
calculate the singular limit of the Lie brackets as $\epsilon \rightarrow 0$.
Physically the velocities are small when compared to the speed of light,
and the spacelike intervals are small when compared to the timelike intervals.
Geometrically we are describing spacetime near a timelike geodesic,
as we are contracting to the subgroup that leaves this worldline invariant,
and so are passing from relativistic to absolute time.  So $adS$
is contracted to $N_-$ while $dS$ is contracted to $N_+$, for example.

{\bfseries \itshape Speed-time contractions.}
We make the substitutions $K \rightarrow \epsilon K$ and $H \rightarrow \epsilon H$
into the Lie algebra and then calculate the singular limit of the Lie brackets
as $\epsilon \rightarrow 0$.  Physically the velocities are small when compared
to the speed of light, and the timelike intervals are small
when compared to the spacelike intervals.
Geometrically we are describing spacetime near
a spacelike geodesic, as we are contracting to the
subgroup that leaves invariant this set of simultaneous events,
and so are passing from relativistic to absolute space.
Such a spacetime may be of limited physical interest,
as we are only considering intervals connecting events that are not causally related.

{\bfseries \itshape Space-time contractions.}
We make the substitutions $P \rightarrow \epsilon P$
and $H \rightarrow \epsilon H$ into the Lie algebra
and then calculate the singular limit of the Lie brackets
as $\epsilon \rightarrow 0$.  Physically the spacelike
and timelike intervals are small, but the boosts are not restricted.
Geometrically we are describing spacetime near an event,
as we are contracting to the subgroup that leaves invariant
only this one event, and so we call the corresponding kinematical
group a {\it local group} as opposed to a {\it cosmological group}.

Fig.~2 illustrates several interesting
relationships among the kinematical groups.
For example, Table~7 gives important classes of kinematical groups,
each class corresponding to a face of the f\/igure, that  transform to
another class in the table under one of the symmetries $S_H$, $S_P$,
or $S_K$, provided that certain exclusions are made as outlined in Table~8.
 The exclusions are necessary under the given symmetries as
 some kinematical algebras are taken to algebras that are not kinematical.

 \begin{figure}[t]
\begin{center}
\includegraphics[width=12.5cm]{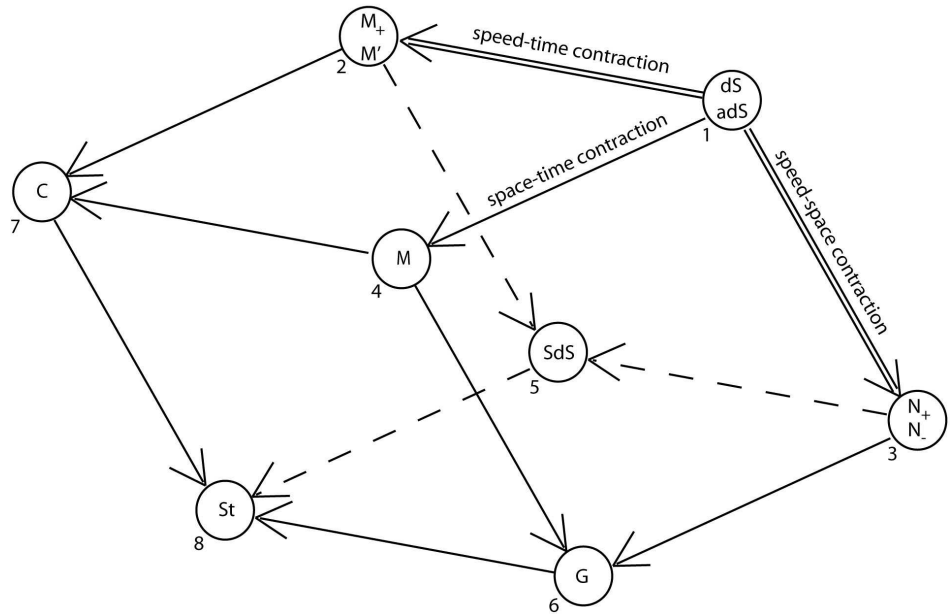}
\end{center}
\caption{The contractions of the kinematical groups for 2-dimensional spacetimes.}
\end{figure}

\begin{table}[t]
 \centering
\caption{Important classes of kinematical groups and their geometrical conf\/igurations in Fig.~2.}
\vspace{1mm}

 \begin{tabular}{ l | l } \hline
 Class of groups & Face \\ \hline \hline
 Relative-time & $1247$ \\
 Absolute-time & $3568$ \\
 Relative-space & $1346$ \\
 Absolute-space & $2578$ \\
 Cosmological & $1235$ \\
 Local & $4678$ \\ \hline
 \end{tabular}
 \end{table}

\begin{table}[t]
 \centering
\caption{The 3 basic symmetries are represented by ref\/lections of Fig.~2, with some exclusions.}

\vspace{1mm}

 \begin{tabular}{ c | c } \hline
 Symmetry & Ref\/lection across face \\ \hline \hline
 $S_H$ & $1378$ (excluding $M_+$) \\
 $S_P$ & $1268$ (excluding $adS$ and $N_-$)\\
 $S_K$ & $1458$ \\ \hline
 \end{tabular}
 \end{table}


\section[Cayley-Klein geometries]{Cayley--Klein geometries}

In this section we wish to review work done by Ballesteros, Herranz, Ortega
and Santander on homogeneous spaces that are spacetimes
for kinematical groups, and we begin with a bit of history concerning
the discovery of non-Euclidean geometries.
Franz Taurinus was the f\/irst to explicitly give mathematical
details on how a hypothetical sphere of imaginary radius would have a
non-Euclidean geometry, what he called {\it log-spherical geometry},
 and this was done via hyperbolic trigonometry (see \cite{G89} or \cite{K98}).
 Felix Klein\footnote{Roger Penrose \cite{P05}
 notes that it was Eugenio Beltrami who f\/irst discovered
 both the projective and conformal models of the hyperbolic plane.}
 is usually given credit for being the f\/irst to give a complete model of
 a non-Euclidean geometry\footnote{ Spherical geometry was not historically considered
 to be non-Euclidean in nature, as it can be embedded in a 3-dimensional Euclidean space,
 unlike Taurinus' sphere.}: he built his model by suitably adapting
 Arthur Cayley's metric for the projective plane.  Klein \cite{K21} (originally published in 1871)
 went on, in a systematic way, to describe nine types of two-dimensional geometries
 (what Yaglom~\cite{Y79} calls {\it Cayley--Klein geometries})
 that were then further investigated by Sommerville~\cite{S10}.
 Yaglom gave conformal models for these geometries, extending what
 had been done for both the projective and hyperbolic planes.
 Each type of geometry is homogeneous and can be determined by
 two real constants $\ka$ and $\kb$ (see Table~9).
The names of the geometries when $\kb \leq 0$ are those as given by Yaglom,
and it is these six geometries that can be interpreted as spacetime geometries.

\begin{table}[t]
 \centering
 \caption{The 9 types of Cayley--Klein geometries.}
 \vspace{1mm}

 \begin{tabular}{ l l l l } \hline
&  \multicolumn{3}{c }{Metric Structure}   \\  \cline{2-4}
 Conformal  & Elliptic & Parabolic & Hyperbolic \\
Structure & $\ka > 0$ & $\ka = 0$ & $\ka < 0$ \\ \hline \hline
\tsep{1ex} Elliptic & elliptic & Euclidean & hyperbolic \\
$\kb > 0$ & geometries & geometries & geometries \\[1mm]
Parabolic & co-Euclidean & Galilean & co-Minkowski  \\
$\kb = 0$ & geometries  & geometry & geometries  \\[1mm]
Hyperbolic & co-hyperbolic & Minkowski & doubly \\
$\kb < 0$ & geometries & geometries & hyperbolic \\
                &                    &                     & geometries \\ \hline
 \end{tabular}
 \end{table}

Following Taurinus, it is easiest to describe a bit of the geometrical
nature of these geometries by applying the appropriate kind of trigonometry:
we will see shortly how to actually construct a model for each geometry.
Let $\kappa$ be a real constant.  The unit circle $a^2 + \kappa b^2 = 1$
in the plane $\R^2 = \{ (a,b) \}$ with metric $ds^2 = da^2 + \kappa db^2$ can be used to def\/ined the cosine
\begin{equation*}
\Cc(\phi) =
\begin{cases}
\cos{\left(\sqrt{\kappa} \, \phi \right) },                   &\text{if $\kappa  > 0$},   \\
1,  &\text{if $\kappa = 0$},      \\
\cosh{\left( \sqrt{-\kappa} \, \phi \right) },   &\text{if $\kappa < 0$},           \\
\end{cases}
\end{equation*}
and sine
\begin{equation*}
\Sc(\phi) =
\begin{cases}
\frac{1}{\sqrt{\kappa}}\sin{\left( \sqrt{\kappa} \, \phi \right) },                   &\text{if $\kappa > 0$},   \\
\phi,  &\text{if $\kappa = 0$},      \\
\frac{1}{\sqrt{-\kappa}} \sinh{\left( \sqrt{-\kappa} \, \phi \right) },   &\text{if $\kappa < 0$}           \\
\end{cases}
\end{equation*}
functions:  here $(a,b) = (\Cc(\phi), \Sc(\phi))$ is a point on
the connected component of the unit circle containing the
point $(1,0)$, and $\phi$ is the signed distance from $(1,0)$
to $(a,b)$ along the circular arc, def\/ined modulo the
length $\dfrac{2\pi}{\sqrt{\kappa}}$ of the unit circle when $\kappa > 0$.
We can also write down the power series for these analytic trigonometric functions:
\begin{gather*}
\Cc(\phi) = 1 - \frac{1}{2!}\kappa \phi^2 + \frac{1}{4!} \kappa^2 \phi^4 + \cdots,
\\
\Sc(\phi) = \phi - \frac{1}{3!}\kappa \phi^3 + \frac{1}{5!} \kappa^2 \phi^5 + \cdots.
\end{gather*}
Note that $\Cc^2(\phi) + \kappa \Sc^2(\phi) = 1$.   So if $\kappa > 0$ then
the unit circle is an ellipse (giving us elliptical trigonometry),
while if $\kappa < 0$ it is a hyperbola (giving us hyperbolic trigonometry).
When $\kappa = 0$ the unit circle consists of two parallel straight lines,
and we will say that our trigonometry is parabolic.
We can use such a trigonometry to def\/ine the angle $\phi$
between two lines, and another independently chosen
trigonometry to def\/ine the distance between two points
(as the angle between two lines, where each line passes
through one of the points as well as a distinguished point).

At this juncture it is not clear that such geometries,
as they have just been described, are of either
mathematical or physical interest.  That mathematicians
and physicists at the beginning of the 20th century
were having similar thoughts is perhaps not surprising,
and Walker \cite{W99} gives an interesting account
of the mathematical and physical research into non-Euclidean
geometries during this period in history.
Klein found that there was a fundamental unity
to these geometries, and so that alone made them worth studying.
Before we return to physics, let us look at these geometries
from a perspective that Klein would have appreciated, describing their motion groups in a unif\/ied manner.

Ballesteros, Herranz, Ortega and Santander have constructed
the Cayley--Klein geometries as homogeneous
spaces\footnote{See \cite{BH06,HOS96,HS02}, and also \cite{HOS00},
where a special case of the group law is investigated,
leading to a plethora of trigonometric identities,
some of which will be put to good use in this paper: see Appendix~A.}
by looking at real representations of their motion groups.
These motion groups are denoted by $\SO$ (that we will refer to as
the {\it generalized} $SO(3)$ or  simply by $SO(3)$)
with their respective Lie algebras being denoted by $\so$
(that we will refer to as the {\it generalized} $so(3)$ or simply by $so(3)$),
and most if not all of these groups are probably familiar to the reader
(for example, if both $\ka$ and $\kb$ vanish, then $SO(3)$ is the Heisenberg group).
Later on in this paper we will use Clif\/ford algebras to show how we can
explicitly think of $SO(3)$ as a rotation group, where each element of $SO(3)$
has a well-def\/ined axis of rotation and rotation angle.

Now a matrix representation of $so(3)$ is given by the matrices
\[
   H =
   \left(
   \begin{matrix}
   0 & -\ka & 0 \\
   1 &     0 & 0 \\
   0 &    0 & 0
   \end{matrix}
   \right), \qquad
      P =
      \left(
      \begin{matrix}
      0 & 0 & -\ka \kb \\
      0 & 0 & 0 \\
      1 & 0 & 0
      \end{matrix}
      \right),  \qquad \mbox{and} \qquad
         K =
         \left(
         \begin{matrix}
         0 & 0 & 0 \\
         0 & 0 & -\kb \\
         0 & 1 & 0
         \end{matrix}
         \right),
 \]
where the structure constants are given by the commutators
\[ \left[ K, H \right] = P,  \qquad \left[ K, P \right]
= -\kb H, \qquad \mbox{and}  \qquad \left[ H, P \right] = \ka K.  \]
By normalizing the constants we obtain matrix representations of the $adS$, $dS$,
$N_-$, $N_+$, $M$, and $G$ Lie algebras, as well as the Lie algebras for the elliptic, Euclidean,
and hyperbolic motion groups, denoted $El$, $Eu$, and $H$ respectively.
We will see at the end of this section how the Cayley--Klein
spaces can also be used to give homogeneous spaces for $M^{\prime}$, $M_+$, $C$,
and $SdS$ (but not for $St$).  One benef\/it of not normalizing
the parameters $\ka$ and $\kb$ is that we can easily obtain contractions
by letting $\ka \rightarrow 0$ or $\kb \rightarrow 0$.

Elements of $SO(3)$ are real-linear, orientation-preserving
isometries of $\R^3 = \{ (z, t, x)) \}$ imbued with the
(possibly indef\/inite or degenerate) metric $ds^2 = dz^2 + \ka dt^2 +  \ka \kb dx^2$.
The one-parameter subgroups $\mathcal{H}$, $\mathcal{P}$,
and $\mathcal{K}$ generated respectively by $H$, $P$, and $K$ consist of matrices of the form
\[
   e^{\alpha H} =
   \left(
   \begin{matrix}
   C_{\ka}(\alpha) & -\ka S_{\ka}(\alpha) & 0 \\
   S_{\ka}(\alpha) & C_{\ka}(\alpha)        & 0 \\
   0                       &  0                             & 1
   \end{matrix}
   \right),
\qquad
   e^{\beta P} =
   \left(
   \begin{matrix}
   C_{\ka \kb}(\beta) & 0 & -\ka \kb S_{\ka \kb}(\beta) \\
   0                           & 1 & 0        \\
   S_{\ka \kb}(\beta) & 0 & C_{\ka \kb}(\beta)
   \end{matrix}
   \right),
\]
\bigskip
and
\[
   e^{\theta K} =
   \left(
   \begin{matrix}
   1 & 0 & 0 \\
   0 & C_{\kb}(\theta)  & -\kb S_{\kb}(\theta)  \\
   0 & S_{\kb}(\theta)  & C_{\kb}(\theta)
   \end{matrix}
   \right)
\]
(note that the orientations induced on the coordinate planes
may be dif\/ferent than expected).   We can now see that in order for $\mathcal{K}$
to be non-compact, we must have that $\kb \leq 0$, which explains the content of Table~3.

The spaces $SO(3) / \mathcal{K}$, $SO(3) / \mathcal{H}$,
and $SO(3) / \mathcal{P}$ are homogeneous spaces for $SO(3)$.
When $SO(3)$ is a kinematical group, then $S \equiv SO(3) / \mathcal{K}$
can be identif\/ied with the manifold of space-time translations.
Regardless of the values of $\ka$ and $\kb$ however, $S$ is the Cayley--Klein
geometry with parameters $\ka$ and $\kb$, and $S$ can be shown to have
constant curvature $\ka$ (also, see \cite{M06}).   So the angle between
two lines passing through the origin (the point that is invariant under
the subgroup $\mathcal{K}$) is given by the parameter $\theta$ of the
element of $\mathcal{K}$ that rotates one line to the other (and
so the measure of angles is related to the parameter $\kb$).
Similarly if one point can be taken to another by an element
of $\mathcal{H}$ or $\mathcal{P}$ respectively, then the distance
between the two points is given by the parameter $\alpha$ or $\beta$,
(and so the measure of distance is related to the parameter $\ka$ or to $\ka \kb$).
Note that the spaces $SO(3) / \mathcal{H}$ and $SO(3) / \mathcal{P}$
are respectively the spaces of timelike and spacelike geodesics for kinematical groups.

For our purposes we will also need to model $S$ as a projective geometry.
First, we def\/ine the projective quadric $\bar{\Sigma}$ as the
set of points on the unit sphere
$\Sigma \equiv \{ (z, t, x) \in \R^3 \; | \; z^2 + \ka t^2 + \ka \kb x^2  = 1 \}$
that have been identif\/ied by the equivalence relation $(z, t, x) \thicksim (-z, -t, -x)$.
The group $SO(3)$ acts on $\bar{\Sigma}$, and the
subgroup $\mathcal{K}$ is then the isotropy subgroup
of the equivalence class $\mathcal{O} = [(1, 0, 0)]$.
The metric $g$ on $\R^3$ induces a metric on $\bar{\Sigma}$ that has $\ka$ as a factor.
If we then def\/ine the main metric $g_1$ on $\bar{\Sigma}$ by setting
\[ \left( ds^2 \right)_1 = \frac{1}{\ka} ds^2, \]
then the surface $\bar{\Sigma}$, along
with its main metric (and subsidiary metric, see below),
is a projective model for the Cayley--Klein geometry $S$.
Note that in general $g_1$ can be indef\/inite as well as nondegenerate.

The motion $\exp(\theta K$) gives a rotation (or boost for a spacetime)
of $S$, whereas the motions $\exp(\alpha H$) and $\exp(\beta P$)
give translations of $S$ (time and space translations respectively for a~spacetime).
The parameters $\ka$ and $\kb$ are, for the spacetimes, identif\/ied
with the universe time radius $\tau$ and speed of light $c$ by the formulae
\[  \ka = \pm \frac{1}{\tau^2} \qquad \mbox{and} \qquad \kb = - \frac{1}{c^2}.  \]

For the absolute-time spacetimes with kinematical groups $N_-$, $G$,
and $N_+$, where $\kb = 0$ and $c = \infty$, we foliate $S$
so that each leaf consists of all points that are simultaneous
with one another, and then $SO(3)$ acts transitively on each leaf.
We then def\/ine the subsidiary metric $g_2$ along each leaf of the foliation by setting
\[
\left( ds^2 \right)_2 = \frac{1}{\kb} \left( ds^2 \right)_1.
\]
Of course when $\kb \neq 0$, the subsidiary metric
can be def\/ined on all of $\bar{\Sigma}$.  The group $SO(3)$ acts
on $S$ by isometries of $g_1$, by isometries of $g_2$ when $\kb \neq 0$
and, when $\kb = 0$,  on the leaves of the foliation by isometries of $g_2$.

It remains to be seen then how homogeneous spacetimes for the
kinematical groups $M_+,$~$M^{\prime},$~$C,\!$ and $SdS$ may be obtained from the Cayley--Klein geometries.
In Fig.~3 the face $1346$ contains the motion groups for all nine types of Cayley--Klein geometries,
and the symmetries $S_H$, $S_P$, and $S_K$ can be represented as
symmetries of the cube, as indicated in
Table~10\footnote{Santander \cite{mS01} discusses
some geometrical consequences of such symmetries when
applied to $dS$, $adS$, and $H$: note that $S_H$, $S_P$, and $S_K$ all f\/ix vertex $1$.}.
As vertices~$1$ and~$8$ are in each of the three planes of ref\/lection,
it is impossible to get $St$ from any one of the Cayley--Klein
groups through the symmetries $S_H$, $S_P$, and $S_K$.
Under the symmetry $S_K$, respective spacetimes for $M_+$, $M^{\prime}$,
and $C$ are given by the spacetimes $SO(3)/\mathcal{K}$ for $N_+$, $N_-$, and $G$,
where space and time translations are interchanged.
\begin{figure}[t]
\begin{center}
\includegraphics[width=12.5cm]{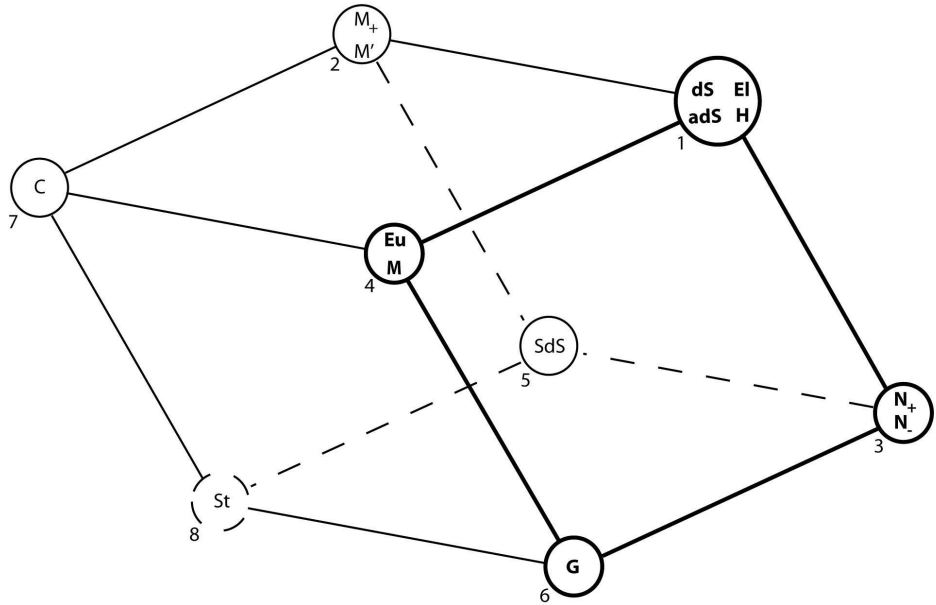}
\end{center}
\caption{The 9 kinematical and 3 non-kinematical groups.}
\end{figure}

\begin{table}[t]
 \centering
\caption{The 3 basic symmetries are given as ref\/lections of Fig.~3.}
\vspace{1mm}
 \begin{tabular}{ c | c } \hline
 Symmetry & Ref\/lection across face \\ \hline \hline
 $S_H$ & $1378$ \\
 $S_P$ & $1268$ \\
 $S_K$ & $1458$ \\ \hline
 \end{tabular}
 \end{table}

Under the symmetry $S_H$, the spacetime
for $SdS$ is given by the homogeneous space $SO(3)/\mathcal{P}$ for $G$,
as boosts and space translations are interchanged by $S_H$.
Note however that there actually are no spacelike geodesics for $G$,
as the Cayley--Klein geometry $S = SO(3)/\mathcal{K}$ for $\ka = \kb = 0$
can be given simply by the plane $\R^2 = \{ (t,x) \}$ with $ds^2 = dt^2$
as its line element\footnote{Yaglom writes in \cite{Y79} about this geometry,
{\it ``\dots which, in spite of its relative simplicity,
confronts the uninitiated reader with many surprising results.''}}.
Although $SO(3)/\mathcal{P}$ is a homogeneous space for $SO(3)$, $SO(3)$
does not act ef\/fectively on $SO(3)/\mathcal{P}$:  since both $[K,P] = 0$ and $[H,P] = 0$,
space translations do not act on $SO(3)/\mathcal{P}$.   Similarly, inertial transformations
do not act on spacetime for $SdS$, or on $St$ for that matter.
Note that $SdS$ can be obtained from $dS$ by $P \rightarrow \epsilon P$,
$H \rightarrow \epsilon H$, and $K \rightarrow \epsilon^2 K$,
where $\epsilon \rightarrow 0$.  So velocities are negligible even when
compared to the reduced space and time translations.

In conclusion to Part I then, a study of all nine types of Cayley--Klein
geometries af\/fords us a beautiful and unif\/ied study of all 11
possible kinematics save one, the static kinematical structure.
It was this study that motivated the author to investigate another
unif\/ied approach to possible kinematics, save for that of the Static Universe.


\pdfbookmark[1]{Part II.  Another unified approach to possible kinematics}{part2}
\section*{Part II.  Another unif\/ied approach to possible kinematics}


\section[The generalized Lie algebra $so(3)$]{The generalized Lie algebra $\boldsymbol{so(3)}$}

Preceding the work of Ballesteros, Herranz, Ortega,
and Santander was the work of Sanjuan~\cite{F84}
on possible kinematics and the nine\footnote{Sanjuan and Yaglom
both tacitly assume that both parameters $\ka$ and $\kb$ are normalized.}
Cayley--Klein geometries.   Sanjuan represents each kinematical Lie algebra as
a real matrix subalgebra of $M(2,\C)$, where $\C$ denotes the generalized
complex numbers (a description of the generalized complex numbers is given below).
This is accomplished using Yaglom's analytic representation of each Caley--Klein
geometry as a region of $\C$: for the hyperbolic plane this gives the
well-known Poincar\'{e} disk model.  Sanjuan
constructs the Lie algebra for the hyperbolic plane using the standard method,
stating that this method can be used to obtain the other Lie algebras as well.
Also, extensive work has been done by Gromov \cite{nG90a,nG90b,nG92,nG95,nG96}
on the generalized orthogonal groups $SO(3)$ (which we refer to simply as $SO(3)$),
deriving representations of the generalized $so(3)$
(which we refer to simply as $so(3)$) by utilizing the dual
numbers as well as the standard complex numbers, where again
it is tacitly assumed that the parameters $\ka$ and $\kb$ have been normalized.
Also, Pimenov has given an axiomatic description of all Cayley--Klein
spaces in arbitrary dimensions in his paper \cite{P65} via the dual numbers $ i_k$, $k=1,2,\dots $,
where $ i_k i_m = i_m i_k \neq 0$ and $i_k^2=0$.

Unless stated otherwise, we will not assume that the parameters
$\ka$ and $\kb$ have been normalized, as we wish to obtain
contractions by simply letting $\ka \rightarrow 0$ or $\kb \rightarrow 0$.
 Our goal in this section is to derive representations of $so(3)$
 as real subalgebras of $M(2,\C)$, and in the process give a
 conformal model of $S$ as a region of the generalized complex plane $\C$
 along with a hermitian metric, extending what has been done for the
 projective and hyperbolic planes\footnote{Fjelstad and Gal \cite{FG01}
 have investigated two-dimensional geometries and physics generated by
 complex numbers from a topological perspective.  Also, see \cite{CCCZ}.}.
 We feel that it is worthwhile to write
down precisely how these representations are obtained in order
that our later construction of a Clif\/ford algebra is more meaningful.

The f\/irst step is to represent the generators
of $SO(3)$ by M\"{o}bius transformations (that is, linear
fractional transformations) of an appropriately def\/ined
region in the complex number plane $\C$, where the points
of $S$ are to be identif\/ied with this region.

\begin{definition}
By the complex number plane $\C_{\kappa}$ we will mean
$\{ w = u + i v \, | \, (u,v) \in \R^2 \ \mbox{and} \ i^2 = -\kappa \}$ where $\kappa$ is a real-valued parameter.
\end{definition}

Thus $\C_{\kappa}$ refers to the complex numbers,
dual numbers, or double numbers when $\kappa$ is
normalized to $1$, $0$, or $-1$ respectively (see \cite{Y79} and \cite{HH04}).
One may check that $\C_{\kappa}$ is an associative algebra with
a multiplicative unit, but that there are zero divisors when $\kappa \leq 0$.
For example, if $\kappa = 0$, then $i$ is a zero-divisor.
The reader will note below that $\frac{1}{i}$ appears in certain
equations, but that these equations can always be rewritten without
the appearance of any zero-divisors in a denominator.
One can extend $\C_{\kappa}$ so that terms like $\frac{1}{i}$
are well-def\/ined (see \cite{Y79}).  It is these zero divisors
that play a crucial rule in determining the null-cone
structure for those Cayley--Klein geometries that are spacetimes.

\begin{definition}
Henceforward $\C$ will denote $\C_{\kb}$, as it is
the parameter $\kb$ which determines the conformal structure
of the Cayley--Klein geometry $S$ with parameters $\ka$ and $\kb$.
\end{definition}

\begin{theorem}
The matrices $\frac{i}{2} \sic$, $\frac{i}{2} \sia$, and $\frac{1}{2i}\sib$
are generators for the generalized Lie algebra $so(3)$,
where $so(3)$ is represented as a subalgebra of the real matrix algebra $M(2,\C)$, where
\[
   \sic =
    \left(
   \begin{matrix}
   1 & 0 \\
   0 &  -1
   \end{matrix}
   \right) , \qquad
   \sia =
    \left(
   \begin{matrix}
   0 & 1 \\
   \ka &  0
   \end{matrix}
   \right) \qquad \mbox{and} \qquad
   \sib =
    \left(
   \begin{matrix}
   0 & i \\
   -\ka i &  0
   \end{matrix}
   \right) .
\]
In fact, we will show that $\mathcal{K}$, $\mathcal{H}$,
and $\mathcal{P}$ (the subgroups generated respectively
by boosts, time and space translations) can be
respectively represented by elements of $SL(2,\C)$
of the form $e^{i\frac{\theta}{2}\sic}$, $e^{i\frac{\alpha}{2}\sia}$, and $e^{\frac{\beta}{2i}\sib}$.
\end{theorem}

 Note that when $\ka =1$ and $\kb = 1$, we recover the
Pauli spin matrices, though my indexing is dif\/ferent,
and there is a sign change as well:  recall that the Pauli spin matrices are typically given as
\[
   \sic =
    \left(
   \begin{matrix}
   0 & 1 \\
   1 & 0
   \end{matrix}
   \right) , \qquad
   \sia =
    \left(
   \begin{matrix}
   0 & -i \\
   i &  0
   \end{matrix}
   \right) \qquad \mbox{and} \qquad
   \sib =
    \left(
   \begin{matrix}
   1 & 0 \\
   0 & -1
   \end{matrix}
   \right) .
\]
We will refer to $\sic$, $\sia$, and $\sib$
as given in the statement of Theorem~1 as the generalized Pauli spin matrices.
\begin{figure}[t]
\begin{center}
\includegraphics[width=8cm]{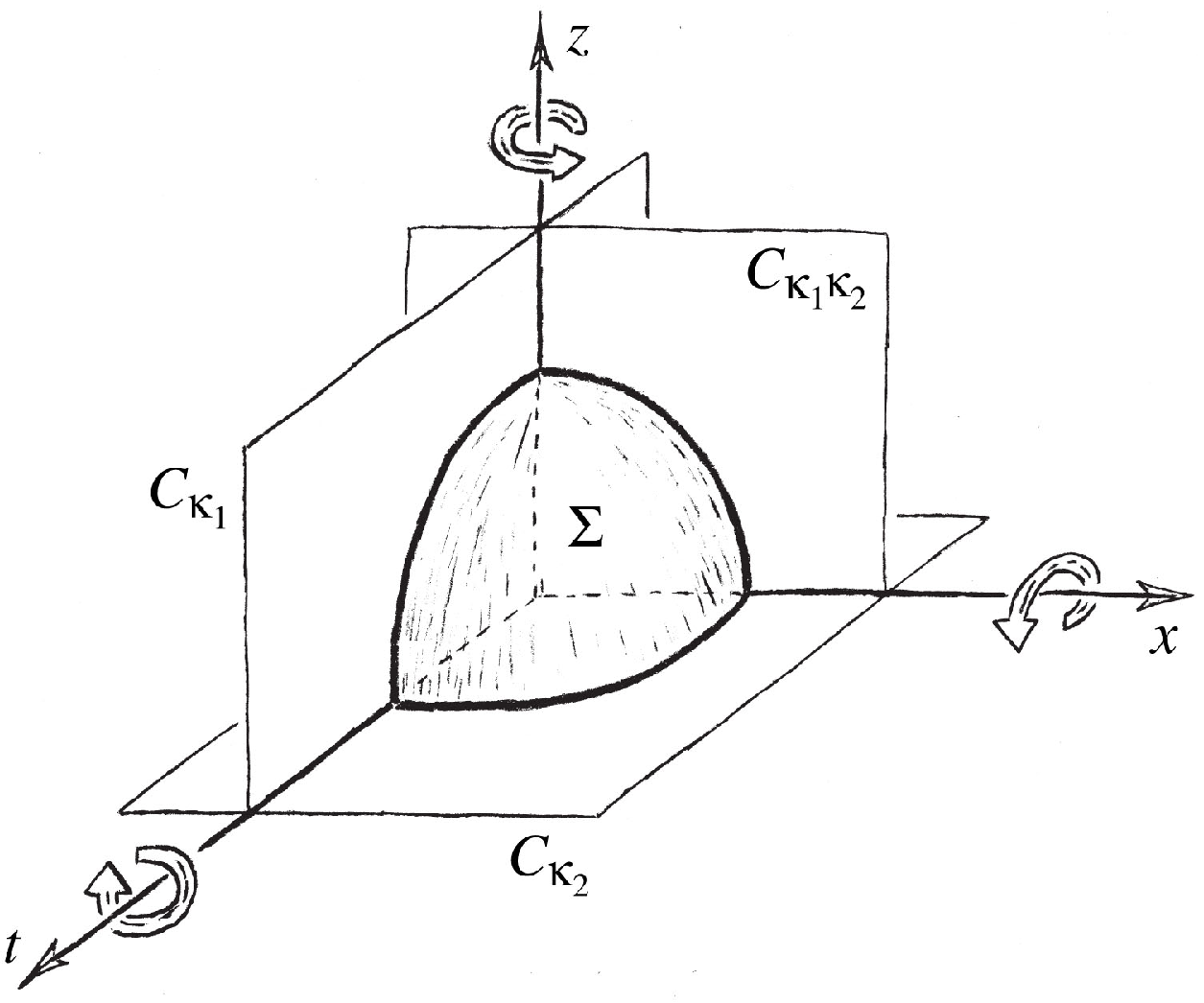}
\end{center}
\caption{The unit sphere $\Sigma$ and the three complex planes $\C_{\kb}$, $\C_{\ka}$, and $\C_{\ka \kb}$.}
\end{figure}

The remainder of this section is devoted to proving the above theorem.
The reader may f\/ind Fig.~4 helpful.
The respective subgroups $\mathcal{K}$, $\mathcal{H}$,
and $\mathcal{P}$ preserve the $z$, $x$, and $t$ axes
as well as the $\C_{\kb}$, $\C_{\ka}$, and $\C_{\ka \kb}$
number planes, acting on these planes as rotations.
Also, as these groups preserve the unit sphere
$\Sigma = \{ (z, t, x) \; | \; z^2 + \ka t^2 + \ka \kb x^2 = 1 \}$,
they preserve the respective intersections of $\Sigma$ with
the $\C_{\kb}$, $\C_{\ka}$, and $\C_{\ka \kb}$ number planes.
These intersections are, respectively, circles
of the form $\ka w\bar{w} = 1$ (there is no intersection
when $\ka = 0$ or when $\ka < 0$ and $\kb > 0$),
$\wa\bar{\wa} = 1$, and $\wb\bar{\wb} = 1$, where $w$, $\wa$,
and $\wb$ denote elements of $\C_{\kb}$, $\C_{\ka}$,
and $\C_{\ka \kb}$ respectively.  We will see in the
next section how a general element of $SO(3)$ behaves
in a~manner similar to the generators of $\mathcal{K}$,
$\mathcal{H}$, and $\mathcal{P}$, utilizing the power of a Clif\/ford algebra.

So we will let the plane $z = 0$ in $\R^3$ represent $\C$
(recall that $\C$ denotes $\C_{\kb}$).  We may then
identify the points of $S$ with a region $\varsigma$
of $\C$ by centrally projecting $\Sigma$ from the point
$(-1, 0, 0)$ onto the plane $z = 0$, projecting only
those points $(z,t,x) \in \Sigma$ with non-negative $z$-values.
The region $\varsigma$ may be open or closed or neither, bounded
or unbounded, depending on the geometry of $S$.
Such a construction is well known for both the
projective and hyperbolic planes ${\bf RP}^2$
and~${\bf H}^2$ and gives rise to the conformal models
of these geometries.   We will see later on how the
conformal structure on $\C$ agrees with that of $S$,
and then how the simple hermitian metric (see Appendix~B)
\[
ds^2 =
\frac{dw d\overline{w}}{\left( 1 + \ka \left| w \right|^2 \right)^2}
\]
gives the main metric $g_1$ for $S$.  This metric can be
used to help indicate the general character of the
region $\varsigma$ for each of the nine types of Cayley--Klein geometries,
as illustrated in Fig.~5.  Note that antipodal points
on the boundary of $\varsigma$ (if there is a boundary)
are to be identif\/ied.  For absolute-time
spacetimes (when $\kb = 0$) the subsidiary metric $g_2$ is given by
\[
g_2 = \frac{dx^2}{\left( 1 + \ka t_0^2 \right)^2}
\]
and is def\/ined on lines $w = t_0$ of simultaneous events.
For all spacetimes, with Here-Now at the origin, the set of zero-divisors gives the null cone for that event.
\begin{figure}[ht]
\begin{center}
\includegraphics[width=12.5cm]{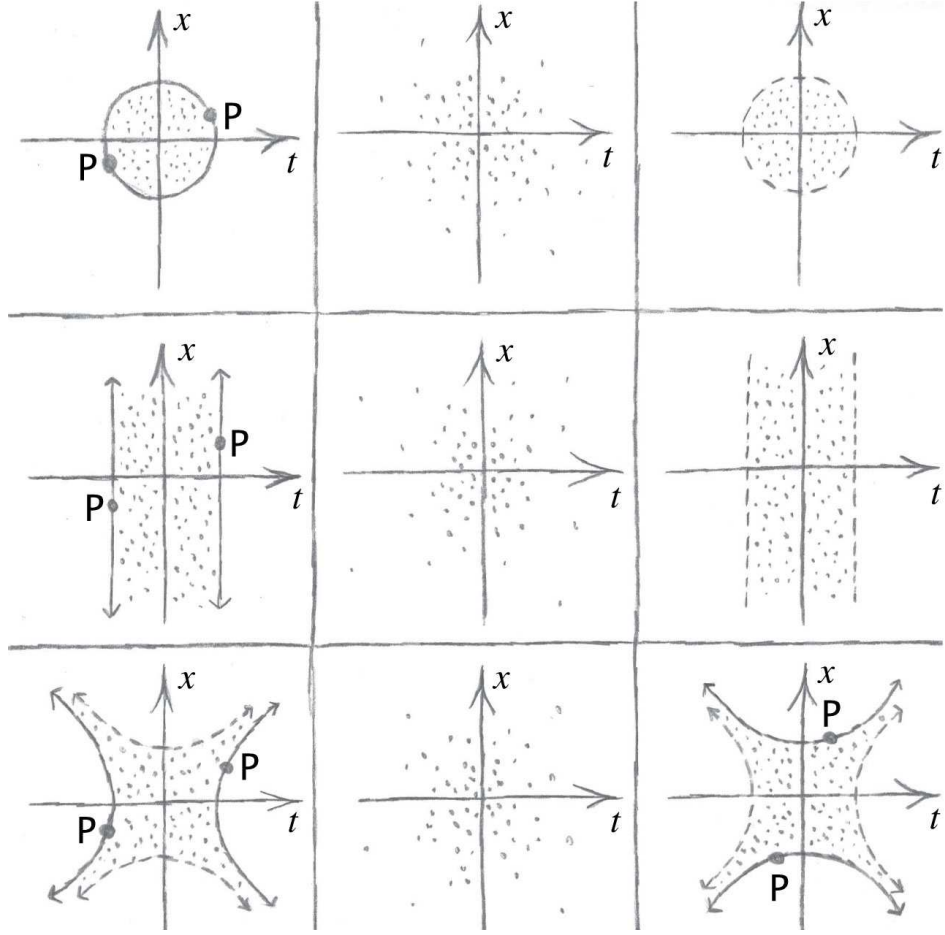}
\end{center}
\caption{The regions $\varsigma$.}
\end{figure}

Via this identif\/ication of points of $S$ with
points of $\varsigma$, transformations of $S$ correspond
to transformations of $\varsigma$.  If the real
parameters $\ka$ and $\kb$ are normalized to the values $\Ka$ and $\Kb$ so that
\begin{equation*}
K_i =
\begin{cases}
1,                   &\text{if $\kappa_i > 0$},   \\
0,  &\text{if $\kappa_i = 0$} ,     \\
-1,   &\text{if $\kappa_i < 0$}
\end{cases}
\end{equation*}
then Yaglom~\cite{Y79} has shown that the linear isometries
of $\R^3$ (with metric $ds^2 = dz^2 + \Ka dt^2 + \Ka\Kb dx^2$)
acting on $\bar{\Sigma}$ project to those M\"{o}bius
transformations that preserve $\varsigma$, and so these M\"{o}bius
transformations preserve
cycles\footnote{Yaglom projects from the point $(z, t, x) =
 (-1, 0, 0)$ onto the plane $z = 1$ whereas we project onto the plane $z = 0$.
 But this hardly matters as cycles are invariant under dilations of $\C$.}:
 a cycle is a curve of constant curvature, corresponding to the
 intersection of a plane in $\R^3$ with $\bar{\Sigma}$.
 We would like to show that elements of $SO(3)$ project to M\"{o}bius
 transformations if the parameters are not normalized, and then to f\/ind
 a realization of $so(3)$ as a real subalgebra of $M(2,\C)$.

Given $\ka$ and $\kb$ we may def\/ine a linear isomorphism of $\R^3$ as indicated below.
\begin{table}[htdp]
\centering
\begin{tabular}{l | l | l} \hline
$\ka \neq 0, \; \kb \neq 0$ & $\ka \neq 0, \; \kb = 0$ & $\ka = \kb = 0$ \\ \hline \hline
$z \mapsto z^{\prime} = z$                                          & $z \mapsto z^{\prime} = z$                                   & $z \mapsto z^{\prime} = z$ \\
$t \mapsto t^{\prime} =  \frac{1}{\sqrt{ | \ka | }} t$        & $t \mapsto t^{\prime}
=  \frac{1    }{\sqrt{ | \ka | }} t$ & $t \mapsto t^{\prime} = t$ \\
\bsep{1ex}$x \mapsto x^{\prime} = \frac{1}{\sqrt{ | \ka \kb | }} x$ & $x \mapsto x^{\prime} = x$
                                 & $x \mapsto x^{\prime} = x$ \\
\hline
\end{tabular}
\end{table}

\noindent
This transformation preserves the projection point $(-1, 0, 0)$
as well as the complex plane $z = 0$, and maps the projective
quadric $\bar{\Sigma}$ for parameters $\Ka$ and $\Kb$ to that
for $\ka$ and $\kb$, and so gives a correspondence between
elements of $SO_{\Ka,\Kb}(3)$ with those of $\SO$ as well
as the projections of these elements.   As the M\"{o}bius
transformations of $\C$ are those transformations that preserve curves of the form
\[
\mbox{Im} \frac{(w^{\prime}_1 - w^{\prime}_3)(w^{\prime}_2
- w^{\prime})}{(w^{\prime}_1 - w^{\prime})(w^{\prime}_2 - w^{\prime}_3)} = 0
\]
(where $w^{\prime}_1$, $w^{\prime}_2$, and $w^{\prime}_3$ are three distinct
points lying on the cycle), then if this form is invariant under the
induced action of the linear isomorphism, then elements of $SO_{\ka,\kb}(3)$
project to M\"{o}bius transformations of $\varsigma$.  As a point $(z, t, x)$
is projected to the point $\left( 0, \frac{t}{z + 1}, \frac{x}{z + 1} \right)$
corresponding to the complex number $w = \frac{1}{z + 1}(t + \mathcal{I}x) \in \C_{\Kb}$,
if the linear transformation sends $(z, t, x)$ to $(z^{\prime}, t^{\prime}, x^{\prime})$,
then it sends $w = \frac{1}{z + 1}(t + \mathcal{I}x) \in \C_{\Kb}$ to $w^{\prime}
= \frac{1}{z^{\prime} + 1}(t^{\prime} + ix^{\prime}) \in \C_{\kb} = \C$,
where $\mathcal{I}^2 = -\Kb$ and $i^2 = -\kb$.  We can then write that
\begin{table}[htdp]
\centering
\begin{tabular}{l | l | l} \hline
$\ka \neq 0, \; \kb \neq 0$ & $\ka \neq 0, \; \kb = 0$ & $\ka = \kb = 0$ \\ \hline \hline
\tsep{1ex} $w = \frac{1}{z+1} \left( t + \mathcal{I}x \right) \mapsto$ &
$w = \frac{1}{z+1}  \left( t + \mathcal{I}x \right) \mapsto$ &
$w = \frac{1}{z+1} \left( t + \mathcal{I}x \right) \mapsto$  \\
$w^{\prime} = \frac{1}{z+1} \frac{1}{\sqrt{| \ka |}} \left( t + \frac{\mathcal{I}}{\sqrt{| \kb |}} x  \right)$ &
$w^{\prime} = \frac{1}{z+1} \left( \frac{1}{\sqrt{| \ka |}} t + \mathcal{I}x  \right)$ &
$ w^{\prime} = w$ \bsep{1ex}\\
\hline
\end{tabular}
\end{table}

\noindent And so
\[
\mbox{Im} \frac{(w_1 - w_3)(w_2 - w)}{(w_1 - w)(w_2 - w_3)} = 0 \iff
\mbox{Im} \frac{(w^{\prime}_1 - w^{\prime}_3)(w^{\prime}_2
- w^{\prime})}{(w^{\prime}_1 - w^{\prime})(w^{\prime}_2 - w^{\prime}_3)} = 0,
\]
as can be checked directly, and we then have that elements of $SO(3)$
project to M\"{o}bius transformations of $\varsigma$.

The rotations $e^{\theta K}$ preserve the complex number plane
$z = t + ix = 0$ and so correspond simply to the transformations of
 $\C$ given by $w \mapsto e^{i \theta}w$, as $e^{i \theta} = \Cb(\theta) + i\Sb(\theta)$,
 keeping in mind that $i^2 = -\kb$.   Now in order to express this rotation as a M\"{o}bius
 transformation, we can write
\[ w \mapsto \frac{e^{\frac{\theta}{2}i}w + 0}{0w + e^{-\frac{\theta}{2}i}} .\]
Since there is a group homomorphism from the subgroup of M\"{o}bius
transformations correspon\-ding to $SO(3)$ to the group $M(2, \C)$ of
 $2 \times 2$ matrices with entries in $\C$, this transformation being def\/ined by
\[  \frac{aw + b}{cw + d} \mapsto
 \left(
   \begin{matrix}
   a & b \\
   c &  d
   \end{matrix}
   \right),
\]
each M\"{o}bius transformation is covered by two elements of $SL(2, \C)$.
So the rotations $e^{\theta K}$ correspond to the matrices
\[
   \pm \left(
   \begin{matrix}
   e^{\frac{\theta}{2}i} & 0 \\
   0 &  e^{-\frac{\theta}{2}i}
   \end{matrix}
   \right) =
  \pm e^{ \frac{\theta}{2}i
    \left(
   \begin{matrix}
   1 & 0 \\
   0 &  -1
   \end{matrix}
   \right) .
   }
\]
For future reference let us now def\/ine {\samepage
\[
   \sic \equiv
    \left(
   \begin{matrix}
   1 & 0 \\
   0 &  -1
   \end{matrix}
   \right) ,
\]
where $\frac{i}{2}\sic$ is then an element of the Lie algebra $so(3)$. }

We now wish to see which elements of $SL(2, \C)$ correspond to the
 motions $e^{\alpha H}$ and $e^{\beta P}$.   The $x$-axis,
 the $zt$-coordinate plane, and the unit sphere $\Sigma$,
 are all preserved by $e^{\alpha H}$.  So the $zt$-coordinate
 plane is given the complex structure $\CA = \{ \wa = z + it \, | \, i^2 = - \ka \}$,
 for then the unit circle $\wa \overline{\wa} = 1$ gives the
 intersection of $\Sigma$ with $\CA$, and the transformation
 induced on $\CA$ by $e^{\alpha H}$ is simply given by $\wa \mapsto e^{i \alpha} \wa$.
 Similarly the transformation induced by $e^{\beta P}$ on
 $\CAB = \{ \wb = z + ix \; | \; i^2 = -\ka \kb \}$ is given by $\wb \mapsto e^{i \beta} \wb$.

In order to explicitly determine the projection of the rotation
$\wa \mapsto e^{i \alpha} \wa$ of the unit circle in~$\CA$ and also
that of the rotation $\wb \mapsto e^{i \beta} \wb$ of the unit circle in $\CAB$,
note that the projection point $(z, t, x) = (-1, 0, 0)$ lies in either
unit circle and that projection sends a point on the unit circle
(save for the projection point itself) to a point on the imaginary axis as follows:
\[ \wa = e^{i \phi}  \mapsto i\Ta\left(\frac{\phi}{2}\right),\qquad
 \wb =  e^{i \phi} \mapsto i\Tab\left(\frac{\phi}{2}\right) \]
(where $T_{\kappa}$ is the tangent function) for
\[  T_{\kappa}\left(\frac{\mu}{2}\right) =   \frac{S_{\kappa}(\mu)}{C_{\kappa}(\mu) + 1}  ,\]
noting that a point $a + ib$ on the unit circle $w \overline{w}$ of the
complex plane $C_{\kappa}$ can be written as $a + ib = e^{i \psi} = \Cc(\psi) + i\Sc(\psi)$.
So the rotations $e^{\alpha H}$ and $e^{\beta P}$ induce the respective transformations
\[ i\Ta\left(\frac{\phi}{2}\right) \mapsto i\Ta\left(\frac{\phi + \alpha}{2}\right),\qquad
i\Tab\left(\frac{\phi}{2}\right) \mapsto i\Tab\left(\frac{\phi + \beta}{2}\right)  \]
on the imaginary axes.  We know that such transformations of either imaginary or
real axes can be extended to M\"{o}bius transformations, and in fact uniquely
determine such M\"{o}bius maps.
For example, if $\wa = i\Ta \left( \frac{\phi}{2} \right)$, then we have that
\[
\wa \mapsto \frac{\wa + i \Ta \left( \frac{\alpha}{2} \right)}
{1 - \frac{ \ka \wa}{i} \Ta \left( \frac{\alpha}{2} \right)}
\]
or
\[
\wa \mapsto \frac{\Ca \left( \frac{\alpha}{2} \right) \wa
+ i \Sa \left( \frac{\alpha}{2} \right)}{-\frac{\ka}{i}
\Sa \left( \frac{\alpha}{2} \right)\wa + \Ca \left( \frac{\alpha}{2} \right)}
\]
with corresponding matrix representation
\[
\pm \left(
\begin{matrix}
\Ca \left( \frac{\alpha}{2} \right) &
i \Sa \left( \frac{\alpha}{2} \right) \vspace{1mm}\\
i \Sa \left( \frac{\alpha}{2} \right) &
\Ca \left(  \frac{\alpha}{2} \right)
\end{matrix}
\right)
\]
in $SL(2, \Ca)$, where we have applied the
trigonometric identity\footnote{For Minkowski spacetimes
this trigonometric identity is the well-known formula for the addition of rapidities.}
\[ T_{\kappa}(\mu \pm \psi) = \frac{T_{\kappa}(\mu) \pm T_{\kappa}(\psi)}
{1 \mp \kappa T_{\kappa}(\mu) T_{\kappa}(\psi)}. \]
However, it is not these M\"{o}bius transformations that we are after,
but those corresponding transformations of $\C$.

Now a transformation of the imaginary axis (the $x$-axis)
of $\CAB$ corresponds to a transformation of the imaginary
axis of $\C$ (also the $x$-axis) while a transformation of
the imaginary axis of $\CA$ (the $t$-axis) corresponds to
a transformation of the real axis of $\C$ (also the $t$-axis).
For this reason, values on the $x$-axis, which are imaginary for
both the $\C_{\ka \kb}$ as well as the $\C$ plane, correspond as
\[
 i \Tab \left( \frac{\phi}{2} \right) = i \frac{1}{\sqrt{\ka}} \Tb \left( \sqrt{\ka} \frac{\phi}{2} \right)
\]
if $\ka > 0$,
\[
 i \Tab \left( \frac{\phi}{2} \right) =
 i \frac{1}{\sqrt{-\ka}} T_{-\kb} \left( \sqrt{-\ka} \frac{\phi}{2} \right)
\]
if $\ka < 0$, and
\[
 i \Tab \left( \frac{\phi}{2} \right) = i  \left( \frac{\phi}{2} \right)
\]
if $\ka = 0$, as can be seen by examining the power series
representation for $T_{\kappa}$.  The situation for the rotation $e^{i \alpha}$ is similar.
We can then compute the elements of $SL(2,\C)$ corresponding
to $e^{\alpha H}$ and $e^{\beta P}$ as given in tables $13$ and $14$
in Appendix~C.  In all cases we have the simple result
that those elements of $SL(2, \C)$ corresponding to $e^{\alpha H}$
can be written as $e^{\frac{\alpha}{2i} \sib}$ and those for $e^{\beta P}$ as $e^{i \frac{\beta}{2} \sia}$, where
\[
   \sia \equiv
    \left(
   \begin{matrix}
   0 & 1 \\
   \ka &  0
   \end{matrix}
   \right) \qquad \mbox{and} \qquad
   \sib \equiv
    \left(
   \begin{matrix}
   0 & i \\
   -\ka i &  0
   \end{matrix}
   \right) .
\]
Thus $\frac{i}{2} \sic$, $\frac{i}{2} \sia$, and $\frac{1}{2i}\sib$
are generators for the generalized Lie algebra $so(3)$, a subalgebra of the real matrix algebra $M(2,\C)$.


\section[The Clifford algebra $Cl_3$]{The Clif\/ford algebra $\boldsymbol{Cl_3}$}

\begin{definition}
Let $Cl_3$ be the 8-dimensional real Clif\/ford algebra that
is identif\/ied with $M(2,\C)$ as indicated by Table~11,
where $\C$ denotes the generalized complex numbers $\C_{\kb}$.
Here we identify the scalar $1$ with the identity matrix and the
volume element $i$ with the $2 \times 2$ identity matrix multiplied
by the complex scalar $i$: in this case $\frac{1}{i}\sib$ can be thought of as the $2 \times 2$ matrix  $\left(
   \begin{matrix}
   0 & 1 \\
   -\ka  &  0
   \end{matrix}
   \right)$.
We will also identify the generalized
Paul spin matrices $\sic$, $\sia$, and $\sib$ with the
vectors $\hat{i} = \langle 1, 0, 0 \rangle$, $\hat{j} = \langle 0, 1, 0 \rangle$,
and $\hat{k} = \langle 0, 0, 1 \rangle$ respectively of
the vector space $\R^3 = \{ (z, t, x) \}$ given the Cayley--Klein
inner product\footnote{We will use the symbol $\hat{v}$ to
denote a vector $v$ of length one under the standard inner product.}.
\end{definition}
\begin{table}[t]
\centering
\caption{The basis elements for $Cl_3$.}
\vspace{1mm}

\begin{tabular}{r c l} \hline
Subspace of & & with basis \\ \hline \hline
scalars & $\R$ & 1 \\
vectors & $\R^3$ & $\sic, \sia, \sib$ \\
bivectors & $\bigwedge^2 \R^3$ & $i \sic, i \sia, \frac{1}{i} \sib$ \\
volume elements & $\bigwedge^3 \R^3$ & $i$ \\
\hline
\end{tabular}
\end{table}

\begin{proposition}
Let $Cl_3$ be the Clifford algebra given by Definition~{\rm 3}.
\begin{itemize}\itemsep=0pt
\item[(i)]  The Clifford product $\sigma_i^2$ gives the square
of the length of the vector $\sigma_i$ under the Cayley--Klein inner product.

\item[(ii)]  The center Cen($Cl_3$) of $Cl_3$ is given by $\R \oplus \bigwedge^3 \R^3$,
the subspace of scalars and volume elements.

\item[(iii)]  The generalized Lie algebra $so(3)$ is isomorphic to the space of bivectors $\bigwedge^2 \R^3$, where
\[
H =  \frac{1}{2i} \sib , \qquad P = \frac{i}{2} \sia , \qquad \mbox{and} \qquad K = \frac{i}{2} \sic .
\]
\item[(iv)]  If $\hat{n} = \langle n^1, n^2, n^3 \rangle$
and $\vec{\sigma} = \langle i\sic, i\sia, \sib/i \rangle$,
then we will let $\ns$ denote the bivector $n^1 i \sic + n^2 i \sia + n^3 \frac{1}{i} \sib$.
This bivector is simple, and the parallel vectors $i\ns$ and $\frac{1}{i} \ns$
are perpendicular to any plane element represented by $\ns$.
Let $\eta$ denote the line through the origin that is determined by $i\ns$ or $\frac{1}{i} \ns$.
\item[(v)] The generalized Lie group $SO(3)$ is also represented within $Cl_3$,
for if $a$ is the vector $a^1\sic + a^2\sia + a^3\sib$, then the linear
transformation of $\R^3$ defined by the inner automorphism
\[
a \mapsto e^{-\frac{\phi}{2} \hat{n} \cdot \vec{\sigma}} a \, e^{\frac{\phi}{2} \hat{n} \cdot \vec{\sigma}}
\]
faithfully represents an element of $SO(3)$ as it preserves vector lengths given by the Cayley--Klein
inner product, and is in fact a rotation, rotating the vector
$\langle a^1, a^2, a^3 \rangle$ about the axis $\eta$ through the angle $\phi$.
In this way we see that the spin group is generated by the elements
\[
 e^{\frac{\theta}{2} i\sic}, \qquad e^{\frac{\beta}{2} i\sia},\qquad
  \mbox{and} \qquad e^{\frac{\alpha}{2i} \sib}  .
\]
\item[(vi)]   Bivectors $\ns$ act as imaginary units
as well as generators of rotations in the oriented planes they represent.
Let $\varkappa$ be the scalar $- \left( \ns \right)^2 $. Then if $a$ lies
in an oriented plane determined by the bivector $\ns$,
where this plane is given the complex structure of $\C_{\varkappa}$,
then $e^{-\frac{\phi}{2} \ns} a  e^{\frac{\phi}{2} \ns}$ is simply
the vector $\langle a^1, a^2, a^3 \rangle$ rotated by the angle $\phi$
in the complex plane $\C_{\varkappa}$, where $\iota^2 = -\varkappa$.
So this rotation is given by unit complex multiplication.
\end{itemize}
\end{proposition}

The goal of this section is to prove Proposition~1.
We can easily compute the following:
\begin{gather*}
\sic^2 = 1, \qquad \sia^2 = \ka , \qquad \sib^2 = \ka \kb ,
\\
\sib\sia  =  -\sia\sib  =  \ka i \sic, \qquad
\sic\sib  =  -\sib\sic  =  i\sia, \\
\sic\sia  =  -\sia\sic  =  \frac{1}{i} \sib ,
\qquad
\sic \sia \sib = -\ka i.
\end{gather*}
Recalling that $\R^3$ is given the Cayley--Klein inner product,
we see that $\sigma_i^2$ gives the square of the length of the vector $\sigma_i$.
Note that when $\ka = 0$, $Cl_3$ is not generated by the vectors.
Cen($Cl_3$) of $Cl_3$ is given by $\R \oplus \bigwedge^3 \R^3$,
and we can check directly that if
\[
H \equiv  \frac{1}{2i} \sib , \qquad
 P \equiv \frac{i}{2} \sia , \qquad \mbox{and} \qquad K \equiv \frac{i}{2} \sic ,
\]
then we have the following commutators:
\begin{gather*}
\left[ H, P \right]     =  HP - PH        =  \frac{1}{4}\left( \sib\sia - \sia\sib \right)    =
\frac{\ka i \sic}{2}  =  \ka K ,\\
\left[ K, H \right]  =  KH - HK  =  \frac{1}{4}\left( \sic\sib - \sib\sic \right)    = \frac{i \sia}{2}       =  P, \\
\left[ K, P \right]  = KP - PK  =  \frac{i^2}{4}\left( \sic\sia - \sia\sic \right)  =
\frac{i \sib}{2}       =  - \kb H.
\end{gather*}
So the Lie algebra $so(3)$ is isomorphic to the space of bivectors $\bigwedge^2 \R^3$.

The product of two vectors $a = a^1\sic + a^2\sia
+ a^3\sib$ and $b = b^1\sic + b^2\sia + b^3\sib$ in $Cl_3$
can be expressed as $ab = a \cdot b + a \wedge b = \frac{1}{2}(ab + ba)
+ \frac{1}{2}(ab - ba)$, where $a \cdot b = \frac{1}{2}(ab + ba)
= a^1 b^1 + \ka a^2 b^2 + \ka \kb a^3 b^3$ is the Cayley--Klein inner product and the wedge product is given by
\[
a \wedge b = \frac{1}{2}(ab - ba) =
\left|
   \begin{matrix}
   -\ka i \sic & -i \sia & \frac{1}{i} \sib \\
   a^1         & a^2    & a^3 \\
   b^1         & b^2    & b^3
   \end{matrix}
   \right|,
\]
so that $ab$ is the sum of a scalar and a bivector: here $| \star |$ denotes the usual $3 \times 3$ determinant.

By the properties of the determinant, if $e \wedge f = g \wedge h$
and $\ka \neq 0$, then the vectors $e$ and $f$
span the same oriented plane as the vectors $g$ and $h$.
When $\ka = 0$ the bivector $\ns$ is no longer simple in the usual way.
For example, for the Galilean kinematical group (aka the Heisenberg group)
where $\ka = 0$ and $\kb = 0$, we have that both $\sic \wedge \sib
= i \sia$ and $\left( \sic + \sia \right) \wedge \sib = i \sia$,
so that the bivector $i \sia$ represents plane elements that do
no all lie in the same plane\footnote{There is some interesting asymmetry for Galilean spacetime,
in that the perpendicular to a timelike geodesic through a given point is uniquely
def\/ined as the lightlike geodesic that passes through that point,
and this lightlike geodesic then has no unique perpendicular, since
all timelike geodesics are perpendicular to it.}.  Recalling that $\sic$, $\sia$,
and $\sib$ correspond to the vectors $\hat{i}$, $\hat{j}$, and $\hat{k}$
respectively, we observe that the subgroup $\mathcal{P}$ of the
Galilean group f\/ixes the $t$-axis and preserves both of these planes,
inducing the same kind of rotation upon each of them:  for the plane spanned by $\hat{i}$ and $\hat{k}$ we have that
\[
 e^{\beta P}:
     \left(
       \begin{matrix}
           \hat{i} \\
           \hat{k}
        \end{matrix}
      \right)
    \mapsto
      \left(
         \begin{matrix}
            \hat{i} + \beta \hat{k} \\
            \hat{k}
         \end{matrix}
       \right)
 \]
 while for the plane spanned by $\hat{i} + \hat{j}$ and $\hat{k}$ we have that
\[
 e^{\beta P}:
     \left(
       \begin{matrix}
           \hat{i} + \hat{j} \\
           \hat{k}
        \end{matrix}
      \right)
    \mapsto
      \left(
         \begin{matrix}
            \hat{i} + \hat{j} + \beta \hat{k} \\
            \hat{k}
         \end{matrix}
       \right).
 \]
If we give either plane the complex structure of the dual
numbers so that $i^2 = 0$, then the rotation is given by
simply multiplying vectors in the plane by the unit complex number $e^{\beta i}$.
We will see below that this kind of construction holds generally.

What we need for our construction below is that any bivector
can be meaningfully expressed as $e \wedge f$ for some vectors $e$ and $f$,
so that the bivector represents at least one plane element: we will discuss
the meaning of the magnitude and orientation of the plane element at the end of the section.
If the bivector represents multiple plane elements spanning distinct planes, so much the better.
If $\hat{n} = \langle n^1, n^2, n^3 \rangle$ and $\vec{\sigma} = \langle i\sic, i\sia, \sib/i \rangle$,
then we will let $\ns$ denote the bivector $B = n^1 i \sic + n^2 i \sia + n^3 \frac{1}{i} \sib$.  Now if
\begin{gather*}
a  = n^1 \sib + \ka n^3 \sic , \qquad
b  = -n^1 \sia + \ka n^2 \sic,\qquad
c  = n^3 \sia + n^2 \sib,
\end{gather*}
then
\begin{gather*}
a \wedge c  = \ka n^3 \ns, \qquad
b \wedge a  = \ka n^1 \ns, \qquad
b \wedge c  = \ka n^2 \ns,
\end{gather*}
where at least one of the bivectors $n^i \ns$ is non-zero as $\ns$ is non-zero.
If $\ka = 0$ and $n^1 = 0$, then $\sic \wedge c = \ns$.
However, if both $\ka = 0$ and $n^1 \neq 0$, then it is impossible
to have $e \wedge f = \ns$:  in this context we may simply replace
the expression $\ns$ with the expression $\sib \wedge \sia$ whenever
$\ka = 0$ and $n^1 \neq 0$ (as we will see at the end of this section,
we could just as well replace $\ns$ with any non-zero multiple of $\sib \wedge \sia$).
The justif\/ication for this is given by letting $\ka \rightarrow 0$, for then
\[
e \wedge f = \left( \sqrt{|\ka|} n^2 \sic - \frac{n^1}{\sqrt{|\ka|}} \sia \right)
\wedge \left( \sqrt{|\ka|} \frac{n^3}{n^1}\sic + \frac{1}{\sqrt{|\ka|}} \sib \right) = \ns
\]
shows that the plane spanned by the vectors $e$ and $f$ tends to the $xt$-coordinate plane.
 We will see below how each bivector $\ns$ corresponds to an element of $SO(3)$
 that preserves any oriented plane corresponding to $\ns$: in the case where
 $\ka = 0$ and $n^1 \neq 0$, we will then have that this element preserves
 the $tx$-coordinate plane, which is all that we require.

It is interesting to note that the parallel vectors $i(a \wedge b)$
and $\frac{1}{i} (a \wedge b)$ (when def\/ined) are perpendicular to both $a$ and $b$
with respect to the Cayley--Klein inner product, as can be checked directly.
However, due to the possible degeneracy of the Cayley--Klein inner product,
there may not be a unique direction that is perpendicular to any given plane.
The vector $i \ns = -\kb n^1 \sic - \kb n^2 \sia + n^3 \sic$ is non-zero
and perpendicular to any plane element corresponding to $\ns$ except when
both $\kb = 0$ and $n^3 = 0$, in which case $i \ns$ is the zero vector.
In this last case the vector $\frac{1}{i} \ns = n^1 \sic + n^2 \sia$ gives a non-zero
normal vector.  In either case, let $\eta$ denote the axis through the origin that
contains either of these normal vectors.

Before we continue, let us reexamine those elements of $SO(3)$
that generate the subgroups $\mathcal{K}$,~$\mathcal{P}$, and $\mathcal{H}$.
Here the respective axes of rotation (parallel to $\sic$, $\sia$,
and $\sib$) for the generators $e^{\theta K}$, $e^{\beta P}$,
and $e^{\alpha H}$ are given by $\eta$, where $\ns$ is given
by $i \sic$ (or $\sib \wedge \sia$ by convention), $i \sia = \sic \wedge \sib$,
and $\frac{1}{i} \sib = \sic \wedge \sia$.
These plane elements are preserved under the respective rotations.
In fact, for each of these planes the rotations are given simply
by multiplication by a~unit complex number, as the $zt$-coordinate
plane is identif\/ied with $\CA$, the $zx$-coordinate plane with $\CAB$,
and the $tx$-coordinate plane with $\CB$ as indicated in Fig.~4.
Note that the basis bivectors act as imaginary units in $Cl_3$ since
\[
\left( \frac{1}{i} \sib \right)^2 = -\ka, \qquad
\left( i \sia \right)^2 = -\ka \kb, \qquad \mbox{and} \qquad
\left( i \sic \right)^2 = -\kb.
\]

The product of a vector $a$ and a bivector $B$ can be written as
$aB = a \dashv B + a \wedge B = \frac{1}{2}(aB - Ba) + \frac{1}{2}(aB + Ba)$
so that $aB$ is the sum of a vector $a \dashv B$ (the left contraction of
 $a$ by~$B$) and a volume element $a \wedge B$.  Let $B = b \wedge c$ for some vectors $b$ and $c$.  Then
\[
2a \dashv (b \wedge c)  = a (b \wedge c) - (b \wedge c)a
                                      = \frac{1}{2} a ( bc - cb) - \frac{1}{2}(bc - cb)a
\]
so that
\begin{gather*}
4a \dashv (b \wedge c)  = cba + abc - acb - bca \\
 \phantom{4a \dashv (b \wedge c)}{} = c(b \cdot a + b \wedge a) + (a \cdot b + a \wedge b)c -
                                       (a \cdot c + a \wedge c)b - b(c \cdot a + c \wedge a) \\
 \phantom{4a \dashv (b \wedge c)}{}  = 2(b \cdot a)c - 2(c \cdot a)b +
                                      c(b \wedge a) + (a \wedge b)c - (a \wedge c)b - b(c \wedge a) \\
\phantom{4a \dashv (b \wedge c)}{} = 2(b \cdot a)c - 2(c \cdot a)b +
                                      c(b \wedge a) - (b \wedge a)c + b(a \wedge c) - (a \wedge c)b \\
\phantom{4a \dashv (b \wedge c)}{} =  2(b \cdot a)c - 2(c \cdot a) b + 2 \left[ c \dashv (b \wedge a) + b \dashv (a
                                             \wedge c) \right] \\
\phantom{4a \dashv (b \wedge c)}{} =  2(b \cdot a)c - 2(c \cdot a) b - 2 a \dashv (c \wedge b) \\
\phantom{4a \dashv (b \wedge c)}{} =  2(b \cdot a)c - 2(c \cdot a) b + 2 a \dashv (b \wedge c)
\end{gather*}
where we have used the Jacobi identity
\[ c \dashv (b \wedge a) + b \dashv (a \wedge c) + a \dashv (c \wedge b) = 0, \]
recalling that $M(2,\C)$ is a matrix algebra where the commutator is given by left contraction.  Thus
\begin{gather*}
2a \dashv (b \wedge c)  = 2(b \cdot a)c - 2(c \cdot a)b
\end{gather*}
and so
\[
a \dashv (b \wedge c) = (a \cdot b) c - (a \cdot c)b.
\]
So the vector $a \dashv B$ lies in the plane determined by the plane element $b \wedge c$.
Because of the possible degeneracy of the Cayley--Klein metric,
it is possible for a non-zero vector $b$ that $b \dashv (b \wedge c) = 0$.

We will show that if $a$ is the vector
$a^1\sic + a^2\sia + a^3\sib$, then the linear transformation of $\R^3$ def\/ined by
\[
a \mapsto e^{-\frac{\phi}{2} \hat{n} \cdot \vec{\sigma}} a \, e^{\frac{\phi}{2} \hat{n} \cdot \vec{\sigma}}
\]
faithfully represents an element of $SO(3)$
(and all elements are thus represented).
In this way we see that the spin group is generated by the elements
\[
 e^{\frac{\theta}{2} i\sic}, e^{\frac{\beta}{2} i\sia}, \qquad \mbox{and} \qquad e^{\frac{\alpha}{2i} \sib}.
\]

First, let us see how, using this construction, the vectors $\sic$, $\sia$,
and $\sib$ (and hence the bivectors $i \sic$, $i \sia$, and $\frac{1}{i} \sib$)
correspond to rotations of the coordinate axes (and hence coordinate planes)
given by $e^{\theta K}$, $e^{\beta P}$, and $e^{\alpha H}$ respectively.
Since
\begin{gather*}
e^{\frac{\theta}{2} i \sic}  = \Cb \left( \frac{\theta}{2} \right) + i \Sb \left( \frac{\theta}{2} \right) \sic,
\qquad
e^{\frac{\beta}{2} i \sia}  = \Cab \left( \frac{\beta}{2} \right) + i \Sab \left( \frac{\beta}{2} \right) \sia, \\
e^{\frac{\alpha}{2i}  \sib}  = \Ca \left( \frac{\alpha}{2} \right) + \frac{1}{i} \Sa \left( \frac{\alpha}{2} \right) \sib
\end{gather*}
and
\begin{gather*}
2\Cc \left( \frac{\phi}{2} \right) \Sc \left( \frac{\phi}{2} \right)                      = \Sc (\phi),\qquad
\Cc^2 \left( \frac{\phi}{2} \right) - \kappa \Sc^2 \left( \frac{\phi}{2} \right)   = \Cc (\phi), \\
\Cc^2 \left( \frac{\phi}{2} \right) + \kappa \Sc^2 \left( \frac{\phi}{2} \right)  = 1
\end{gather*}
(noting that $\Cc$ is an even function while $\Sc$ is odd) it follows that
\begin{gather*}
e^{-\frac{\theta}{2} i \sic} \sigma_j e^{\frac{\theta}{2} i \sic}              =
\begin{cases}
\sic                                                             & \text{if $j = 1$}, \\
\Cb (\theta) \sia - \Sb (\theta) \sib & \text{if $j = 2$}, \\
\Cb (\theta) \sib + \kb \Sb (\theta) \sia            & \text{if $j = 3$},
\end{cases} \\
e^{-\frac{\beta}{2} i \sia} \sigma_j e^{\frac{\beta}{2} i \sia}              =
\begin{cases}
\Cab (\beta) \sic + \Sab (\beta) \sib & \text{if $j = 1$}, \\
\sia                                                             & \text{if $j = 2$}, \\
\Cab (\beta) \sib - \ka \kb \Sab (\beta) \sic            & \text{if $j = 3$},
\end{cases} \\
e^{-\frac{\alpha}{2i}  \sib} \sigma_j e^{\frac{\alpha}{2i}  \sib}           =
\begin{cases}
\Ca (\alpha) \sic + \Sa (\alpha) \sia       & \text{if $j = 1$}, \\
\Ca (\alpha) \sia - \ka \Sa (\alpha) \sic            & \text{if $j = 2$}, \\
\sib                                                             & \text{if $j = 3$}.
\end{cases}
\end{gather*}
So for each plane element, the $\sigma_j$ transform as
the components of a vector under rotation in the clockwise direction,
given the orientations of the respective plane elements:
\[
i\sic \; \mbox{is represented by} \; \sib \wedge \sia, \qquad
 i\sia = \sic \wedge \sib, \qquad \mbox{and} \qquad \frac{1}{i} \sib = \sic \wedge \sia.
\]
Now we can write
\[
e^{\frac{\phi}{2} \ns} =
1 +
\frac{\phi}{2} \ns +
\frac{1}{2!} \left( \frac{\phi}{2} \right)^2 \left( \ns \right)^2 +
\frac{1}{3!} \left( \frac{\phi}{2} \right)^3 \left( \ns \right)^3 + \cdots.
\]
If $\varkappa$ is the scalar $-\left( \ns \right)^2$, then
\begin{gather*}
e^{\frac{\phi}{2} \ns} =  \left(
1 -
\frac{1}{2!} \left( \frac{\phi}{2} \right)^2 \varkappa +
\frac{1}{4!} \left( \frac{\phi}{2} \right)^4 \varkappa^2 - \cdots
\right)\\
\phantom{e^{\frac{\phi}{2} \ns}=}{}
 +  \ns \left(
\frac{\phi}{2} -
\frac{1}{3!} \left( \frac{\phi}{2} \right)^3 \varkappa +
\frac{1}{5!} \left( \frac{\phi}{2} \right)^5 \varkappa^2 - \cdots
\right) \\
\phantom{e^{\frac{\phi}{2} \ns}}{}  =  C_{\varkappa} \left( \frac{\phi}{2} \right)
+  \ns S_{\varkappa} \left( \frac{\phi}{2} \right).
\end{gather*}

As $a = a^1 \sic + a^2 \sia + a^3 \sib$ is a vector, we can compute its length easily
using Clif\/ford multiplication as $a a = (a^1)^2 + \ka (a^2)^2 + \ka \kb (a^3)^2 = | a |^2$.
 We would like to show that $e^{-\frac{\phi}{2} \ns} a  e^{\frac{\phi}{2} \ns}$
 is also a vector with the same length as $a$.   If $g$ and $h$ are elements of a matrix Lie algebra,
 then so is $e^{-\phi \, {\mbox{\tiny ad}} \, g} h = e^{-\phi g} h e^{\phi g}$ (see \cite{SW86} for example).
 So if $B$ is a bivector $B^1 i\sic + B^2 i\sia + B^3\frac{1}{i}\sib$,
 then $e^{-\frac{\phi}{2} \ns} B  e^{\frac{\phi}{2} \ns}$ is also a bivector.
 It follows that $e^{-\frac{\phi}{2} \ns} a  e^{\frac{\phi}{2} \ns}$ is a vector
 as the volume element $i$ lies in Cen($Cl_3$) so that $e^{-\frac{\phi}{2} \ns}
 \sic  e^{\frac{\phi}{2} \ns}$, $e^{-\frac{\phi}{2} \ns} \sia  e^{\frac{\phi}{2} \ns}$,
 and $e^{-\frac{\phi}{2} \ns} \sib  e^{\frac{\phi}{2} \ns}$ are all vectors.  Since
 \[
\left(
e^{-\frac{\phi}{2} \ns} a  e^{\frac{\phi}{2} \ns}
\right)
\left(
e^{-\frac{\phi}{2} \ns} a  e^{\frac{\phi}{2} \ns}
\right) =
e^{-\frac{\phi}{2} \ns} | a |^2  e^{\frac{\phi}{2} \ns}  =
| a |^2  e^{-\frac{\phi}{2} \ns}  e^{\frac{\phi}{2} \ns}  =
\left| a \right|^2
\]
it follows that $e^{-\frac{\phi}{2} \ns} a  e^{\frac{\phi}{2} \ns}$
has the same length as $a$.  So the inner automorphism of $\R^3$ given
by $a \mapsto e^{-\frac{\phi}{2} \ns} a  e^{\frac{\phi}{2} \ns}$ corresponds
to an element of $SO(3)$.  We will see in the next section that all elements of $SO(3)$
are represented by such inner automorphisms of $\R^3$.

Finally, note that $e^{-\frac{\phi}{2} \ns} \left( \ns \right) e^{\frac{\phi}{2} \ns} = \ns$
as $\ns$ commutes with $e^{\frac{\phi}{2} \ns}$: so any plane element
represented by $\ns$ is preserved by the corresponding element of $SO(3)$.
In fact, if $\ns = a \wedge b$ for some vectors $a$ and $b$ and $\varkappa$
is the scalar $ - \left( a \wedge b \right)^2$, then
\begin{gather*}
     e^{-\frac{\phi}{2} a \wedge b} ( a )  e^{\frac{\phi}{2} a \wedge b}
=  \left[ C_{\varkappa} \left( \frac{\phi}{2} \right) -  (a \wedge b )S_{\varkappa}
\left( \frac{\phi}{2} \right) \right] ( a )
          \left[ C_{\varkappa} \left( \frac{\phi}{2} \right)
          +  (a \wedge b) S_{\varkappa} \left( \frac{\phi}{2} \right) \right] \\
\phantom{e^{-\frac{\phi}{2} a \wedge b} ( a )  e^{\frac{\phi}{2} a \wedge b}}{} =  C^2_{\varkappa} \left( \frac{\phi}{2} \right) a
          + C_{\varkappa} \left( \frac{\phi}{2} \right) S_{\varkappa}
          \left( \frac{\phi}{2} \right)  (a \wedge b) a (a \wedge b) \\
\phantom{e^{-\frac{\phi}{2} a \wedge b} ( a )  e^{\frac{\phi}{2} a \wedge b}=}{}
-  C_{\varkappa} \left( \frac{\phi}{2} \right) S_{\varkappa} \left( \frac{\phi}{2} \right) (a \wedge b) a
-       S^2_{\varkappa} \left( \frac{\phi}{2} \right)  a(a \wedge b).
\end{gather*}
Since $a (a \wedge b) = -(a \wedge b) a$, then
\[
e^{-\frac{\phi}{2} a \wedge b} ( a )  e^{\frac{\phi}{2} a \wedge b} =
\left[ C_{\varkappa}(\phi) - (a \wedge b) S_{\varkappa}(\phi) \right] a,
\]
and so vectors lying in the plane determined by $a \wedge b$ are simply
rotated by an angle $-\phi$, and this rotation is given by simple multiplication
by a unit complex number $e^{-i \phi}$ where $i^2 = -\varkappa$.  Thus, the
linear combination $ua + vb$ is sent to $ue^{-i \phi}a + ve^{-i \phi}b$,
and so the plane spanned by the vectors $a$ and $b$ is preserved.

The signif\/icance is that if $a$ lies in an oriented plane determined by
the bivector $\ns$ where this plane is given the complex structure of
$\C_{\varkappa}$, then $e^{-\frac{\phi}{2} \ns} a  e^{\frac{\phi}{2} \ns}$
is simply the vector $a$ rotated by an angle of $-\phi$ in the complex plane
$\C_{\varkappa}$, where $\iota^2 = -\varkappa$.  Furthermore, the axis of
rotation is given by $\eta$ as $\eta$ is preserved (recall that $i$ lies
in the center of $Cl_3$).  Since the covariant components $\sigma_i$ of $a$
are rotated clockwise, the contravariant components $a^j$ are rotated counterclockwise.
So $\langle a^1, a^2, a^3 \rangle$ is rotated by the angle $\phi$ in the complex plane
$C_{\varkappa}$ determined by $\ns$.

If we use $b \wedge a$ instead of $a \wedge b$ to represent the plane element,
then $\varkappa$ remains unchanged.  Note however that, if $c$ is a vector lying in this plane, then
\begin{gather*}
e^{-\frac{\phi}{2} b \wedge a} c e^{-\frac{\phi}{2} b \wedge a}
    = \left[ C_{\varkappa}(\phi) - (b \wedge a) S_{\varkappa}(\phi) \right] c
    = \left[ C_{\varkappa}(-\phi) - (a \wedge b) S_{\varkappa}(-\phi) \right] c
\end{gather*}
so that rotation by an angle of $\phi$ in the plane oriented according
to $b \wedge a$ corresponds to a~rotation of angle $-\phi$
in the same plane under the opposite orientation as given by $a \wedge b$.

It would be appropriate at this point to note two things:
one, the magnitude of $\ns$ appears to be important, since
$\varkappa = - \left( \ns \right)^2$, and two, the normalization
$(n^1)^2 + (n^2)^2 + (n^3)^2 = 1$ of $\hat{n}$ is somewhat
arbitrary\footnote{Due to dimension requirements some kind of
normalization is needed as we cannot have $\phi$, $n^1$, $n^2$, and $n^3$
as independent variables, for $so(3)$ is 3-dimensional.}.
These two matters are one and the same.  We have chosen this normalization
because it is a simple and natural choice.  This particular normalization
is not essential, however.  For suppose that $\varkappa = - (a \wedge b)^2$
while $\varkappa^{\prime} = - (n a \wedge b)^2$, where $n$ is a positive constant.
 Let $\C_{\varkappa} = \{ t + ix \, | \, i^2 = -\varkappa \}$
 with angle measure $\phi$ and $\C_{\varkappa^{\prime}} = \{ t + \iota x \, | \,\iota^2
 = -\varkappa^{\prime} = -n^2 \varkappa \}$ with angle measure $\theta$:
  without loss of generality let $\varkappa > 0$.  Then $\phi = n\theta$, for
\begin{gather*}       e^{i \theta}
 = \cos{\left( \sqrt{\varkappa^{\prime}} \theta \right)}
 - \frac{\iota}{\sqrt{\varkappa^{\prime}}}\sin{\left( \sqrt{\varkappa^{\prime}} \theta \right)}
  = \cos{\left( n\sqrt{\varkappa} \theta \right)} - \frac{\iota}{n\sqrt{\varkappa}}
  \sin{\left( n\sqrt{\varkappa} \theta \right)} \\
 \phantom{e^{i \theta}}{} = \cos{\left( \sqrt{\varkappa} \phi \right)} - \frac{i}{\sqrt{\varkappa}}\sin{\left( \sqrt{\varkappa} \phi \right)}
  = e^{i \phi}.
\end{gather*}
So we see that $SO(3)$ is truly a rotation group, where each
element has a distinct axis of rotation as well as a well-def\/ined rotation angle.


\section[$SU(2)$]{$\boldsymbol{SU(2)}$}

Since the generators of the generalized Lie group $SO(3)$
can be represented by inner automorphisms of the subspace $\R^3$
of vectors of $Cl_3$ (see Def\/inition~3), then every element of $SO(3)$
can be represented by an inner automorphism, as the composition of
inner automorphisms is an inner automorphism.  On the other hand,
we've seen that any inner automorphism represents an element of $SO(3)$.
In fact, each rotation belonging to $SO(3)$ is then represented by two elements
$\pm e^{\frac{\phi}{2} \ns}$ of $SL(2,\C)$, where as usual $\C$ denotes the
generalized complex number $\C_{\kb}$: we will denote the subgroup of $SL(2,\C)$
consisting of elements of the form $\pm e^{\frac{\phi}{2} \ns}$ by $SU(2)$.

\begin{definition}
Let $A$ be the matrix
\[
A =
\left(
   \begin{matrix}
   \ka & 0 \\
   0    & 1
   \end{matrix}
\right) .
\]
\end{definition}

We will now use Def\/inition 4 to show that $SU(2)$ is a subgroup of the subgroup $G$ of $SL(2,\C)$
consisting of those matrices $U$ where $U^{\star} A U = A$: in fact, both these subgroups of
$SL(2,\C)$ are one and the same, as we shall see.
Now
\begin{gather*}
\big( e^{\frac{\phi}{2} \ns}\big)^{\star} A e^{\frac{\phi}{2} \ns}
= \left[ C_{\varkappa}\left( \frac{\phi}{2} \right) +
\left( {\ns} \right)^{\star} S_{\varkappa}\left( \frac{\phi}{2} \right)  \right] A
          \left[ C_{\varkappa}\left( \frac{\phi}{2} \right) + \ns S_{\varkappa}\left( \frac{\phi}{2} \right) \right] \\
 \phantom{\big( e^{\frac{\phi}{2} \ns}\big)^{\star} A e^{\frac{\phi}{2} \ns}}{}
  = C_{\varkappa}^2\left( \frac{\phi}{2} \right)A +
\left( {\ns} \right)^{\star} A \left( \ns \right) S_{\varkappa}^2\left( \frac{\phi}{2} \right) \\
\phantom{\big( e^{\frac{\phi}{2} \ns}\big)^{\star} A e^{\frac{\phi}{2} \ns}=}{}
+ A \left( \ns \right) C_{\varkappa}\left( \frac{\phi}{2} \right)S_{\varkappa}\left( \frac{\phi}{2} \right) +
\left( {\ns} \right)^{\star} A C_{\varkappa}\left( \frac{\phi}{2} \right)S_{\varkappa}\left( \frac{\phi}{2} \right)
 = A
\end{gather*}
because $A \left( \ns \right) = -\left( {\ns} \right)^{\star} A$ implies that
\[
A \left( \ns \right) C_{\varkappa}\left( \frac{\phi}{2} \right)S_{\varkappa}\left( \frac{\phi}{2} \right) +
\left( {\ns} \right)^{\star} A C_{\varkappa}\left( \frac{\phi}{2} \right)S_{\varkappa}\left( \frac{\phi}{2} \right)
= 0
\]
and $\left( {\ns} \right)^{\star} A \left( \ns \right) = -A \left( \ns \right)^2 = \varkappa A$ implies that
\begin{gather*}
 C_{\varkappa}^2\left( \frac{\phi}{2} \right)A +
\left( {\ns} \right)^{\star} A \left( \ns \right) S_{\varkappa}^2\left( \frac{\phi}{2} \right)
 = C_{\varkappa}^2\left( \frac{\phi}{2} \right)A +
\varkappa S_{\varkappa}^2\left( \frac{\phi}{2} \right) A   = A.
\end{gather*}
So $SU(2)$ is a subgroup of the subgroup $G$ of $SL(2,\C)$ consisting of those matrices $U$ where $U^{\star} A U = A$.

We can characterize this subgroup $G$ as
\[
\left\{
\left(
   \begin{matrix}
   \alpha               & \beta \\
   -\ka \overline{\beta} & \overline{\alpha}
   \end{matrix}
\right)
 | \,
\alpha, \beta \in \C \; \mbox{and} \; \alpha \overline{\alpha} + \ka \beta \overline{\beta} = 1
\right\}.
\]
Now
\[
e^{\frac{\phi}{2} \ns} =
\left(
   \begin{matrix}
   C_{\varkappa}^2\left( \frac{\phi}{2} \right) + n^1 i S_{\varkappa}^2\left( \frac{\phi}{2} \right) &
   n^2 i S_{\varkappa}^2\left( \frac{\phi}{2} \right) + n^3 S_{\varkappa}^2\left( \frac{\phi}{2} \right)\vspace{1mm} \\
   n^2 \ka i S_{\varkappa}^2\left( \frac{\phi}{2} \right) - n^3 \ka S_{\varkappa}^2\left( \frac{\phi}{2} \right) &
   C_{\varkappa}^2\left( \frac{\phi}{2} \right) - n^1 i S_{\varkappa}^2\left( \frac{\phi}{2} \right)
   \end{matrix}
\right)
\]
as can be checked directly, recalling that
\[
e^{\frac{\phi}{2} \ns} = C_{\varkappa}\left( \frac{\phi}{2} \right)
+ \left( \ns \right) S_{\varkappa}\left( \frac{\phi}{2} \right),
\]
where
\[
\varkappa = - \left( \ns \right)^2 = \left(n^1\right)^2 \kb
+ \left( n^2 \right)^2 \ka \kb + \left( n^3 \right)^2 \ka.
\]
Thus $\det\left( e^{\frac{\phi}{2} \ns} \right) = 1$,
and we see that any element of $G$ can be written in the form $e^{\frac{\phi}{2} \ns}$.
So the group $SU(2)$ can be characterized by
\[
SU(2) =
\left\{
\left(
   \begin{matrix}
   \alpha               & \beta \\
   -\ka \overline{\beta} & \overline{\alpha}
   \end{matrix}
\right)
 | \,
\alpha, \beta \in \C \; \mbox{and} \; \alpha \overline{\alpha} + \ka \beta \overline{\beta} = 1
\right\}.
\]
Note that if $U(\lambda)$ is a curve passing through the identity at $\lambda = 0$, then
\[
\left. \frac{d}{d\lambda} \right|_{\lambda = 0} \left( U^{\star} A U = A \right) \ \Longrightarrow \
  \dot{U}^{\star}A + A\dot{U} = 0
\]
so that $su(2)$ consists of those elements $B$ of $M(2,\C)$ such that $B^{\star}A + AB = 0$.  Although $SU(2)$ is a double cover of $SO(3)$, it is not necessarily the universal cover for $SO(3)$, nor even connected, for sometimes $SO(3)$ is itself simply-connected.  Thus we have shown that:

\begin{theorem}
The Clifford algebra $Cl_3$ can be used to construct a double cover of the generalized Lie group $SO(3)$,
for a vector $a$ can be rotated by the inner automorphism
\[
\R^3 \rightarrow \R^3, \qquad a \mapsto \mathit{s}^{-1} a \mathit{s}
\]
where $\mathit{s}$ is an element of the group
\[
{\bf Spin}(3) =
\left\{
\left(
   \begin{matrix}
   \alpha               & \beta \\
   -\ka \overline{\beta} & \overline{\alpha}
   \end{matrix}
\right)
 | \,
\alpha, \beta \in \C \; \mbox{and} \; \alpha \overline{\alpha} + \ka \beta \overline{\beta} = 1
\right\},
\]
where $\C$ denotes the generalized complex number $\C_{\kb}$.
\end{theorem}

\begin{lemma}
We define the generalized special unitary group $SU(2)$ to be ${\bf Spin}(3)$.  Then $su(2)$
consists of those matrices $B$ of $M(2,\C)$ such that $B^{\star}A + AB = 0$.
\end{lemma}


\section[The conformal completion of $S$]{The conformal completion of $\boldsymbol{S}$}

Yaglom~\cite{Y79} has shown how the complex plane $\C_{\kappa}$ may be extended
to a Riemann sphere $\Gamma$ or inversive plane\footnote{Yaglom did
this when $\kappa \in \{-1, 0, 1 \}$, but it is a simple matter to generalize his results.}
(and so dividing by zero-divisors is allowed), upon which the entire set
of M\"{o}bius transformations acts globally and so gives a group of conformal transformations.
In this last section we would like to take advantage of the simple structure of
this conformal group and give the conformal completion of $S$, where $S$ is conformally
embedded simply by inclusion of the region $\varsigma$ lying in $\C$ and therefore lying in~$\Gamma$.
Herranz and Santander~\cite{HS02b} found a conformal
completion of $S$ by realizing the conformal group as a group
of linear transformations acting on $\R^4$, and then constructing the
conformal completion as a homogeneous phase space of this conformal group.
The original Cayley--Klein geometry $S$ was then embedded
into its conformal completion by one of two methods, one a group-theoretical
one involving one-parameter subgroups and the other stereographic projection.

The 6-dimensional real Lie algebra for $SL(2,\C)$ consist of those matrices in $M(2,\C)$
with trace equal to zero.  In addition to the three generators $H$, $P$, and $K$
\[
H = \frac{1}{2i}\sib =
\left(
   \begin{matrix}
   0 & \frac{1}{2} \\
   -\frac{\ka}{2} & 0
   \end{matrix}
\right), \qquad
P = \frac{i}{2}\sia =
\left(
   \begin{matrix}
   0 & \frac{i}{2} \\
   \frac{\ka i}{2} & 0
   \end{matrix}
\right), \qquad
K = \frac{i}{2}\sic =
\left(
   \begin{matrix}
   \frac{i}{2} & 0 \\
   0 & -\frac{i}{2}
   \end{matrix}
\right)
\]
that come from the generalized Lie group $SO(3)$
of isometries of $S$, we have three other generators for $SL(2, \C)$:
one, labeled $D$, for the subgroup of dilations centered at the origin and
two others, labeled $G_1$ and $G_2$, for ``translations''.
 It is these transformations $D$, $G_1$, $G_2$, that necessitate extending $\varsigma$
 to the entire Riemann sphere $\Gamma$, upon which the set of M\"{o}bius
 transformations acts as a conformal group.  Note that the following correspondences
 for the M\"{o}bius transformations $w \mapsto w + t$ and $w \mapsto w + ti$
 (for real parameter $t$) are valid only if $\ka \neq 0$, which explains why our
 ``translations" $G_1$ and $G_2$ are not actually translations:
\begin{gather*}
\exp \left[ t \left(
   \begin{matrix}
   0 & 1 \\
   0 & 0
   \end{matrix}
   \right) \right] =
   \left(
   \begin{matrix}
   1 & t \\
   0 & 1
   \end{matrix}
   \right)
   \hspace{0.25in} \rightleftarrows \hspace{0.25in}
   w \mapsto w + t,
\\
\mbox{exp} \left[ t \left(
   \begin{matrix}
   0 & i \\
   0 & 0
   \end{matrix}
   \right) \right] =
   \left(
   \begin{matrix}
   1 & ti \\
   0 & 1
   \end{matrix}
   \right)
   \hspace{0.25in} \rightleftarrows \hspace{0.25in}
   w \mapsto w + ti.
\end{gather*}
Please see Tables 15 and 16.
The structure constants $[\star, \star \star]$ for this basis of $sl(2,\C)$
(which is the same basis as that given in \cite{HS02} save for a sign change in $G_2$) are given by Table~12.

\begin{table}[t]
 \centering
\caption{Additional basis elements for $sl(2,\C)$.}
\vspace{1mm}
 \begin{tabular}{ c ||  c   c   c   c   c   c   c  } \hline
$\star \diagdown \star\star $ & $H$ & $P$ & $K$ & $G_1$ & $G_2$ & $D$  \\ \hline \hline
$H$     & 0                         & $\ka K$        & $-P$        &  $D$         & $K$     & $-H - \ka G_1$  \\
$P$     & $-\ka K$      & 0                          & $\kb H$    & $K$  & $-\kb D$     &  $-P + \ka G_2$ \\
$K$ & $P$                 &  $-\kb H$         & 0                   & $-S_2$    & $\kb G_2$   & $0$                        \\
$G_1$    &  $-D$                   &  $-K$           &  $S_2$          & 0             & $0$              & $G_1$                   \\
$G_2$    &  $-K$           &  $\kb D$               &  $-\kb G_2$  &  $0$         & 0                 & $G_2$                   \\
$D$        & $H + \ka G_1$ &  $P - \ka G_2$ &   $0$             &  $-G_1$   &  $-G_2$      & 0                            \\
\hline
 \end{tabular}
 \end{table}


\appendix

\section{Appendix:  Trigonometric identities}

The following trigonometric identities are taken from \cite{HOS00} and \cite{HS02},
and are used throughout Sections 3, 4, and 5
\begin{gather*}
 \frac{d}{d \phi} C_{\kappa} (\phi) = -\kappa S_{\kappa}(\phi),\qquad
 \frac{d}{d \phi} S_{\kappa} (\phi) = C_{\kappa}(\phi),\qquad
 \frac{d}{d \phi} T^{-1}_{\kappa} (\phi) = \frac{1}{1 + \kappa \phi^2},\\
 C^2_{\kappa} (\phi) + \kappa S^2_{\kappa}(\phi) = 1, \qquad
 C_{\kappa}(2\phi) = C^2_{\kappa} (\phi) - \kappa S^2_{\kappa}(\phi),\qquad
 S_{\kappa} (2\phi) = 2C_{\kappa} (\phi) S_{\kappa}(\phi),\\
 T_{\kappa}\left( \frac{\phi}{2} \right) = \frac{S_{\kappa}(\phi)}{C_{\kappa}(\phi) + 1},\qquad
 T_{\kappa}(\phi \pm \psi) = \frac{T_{\kappa}(\phi)
 \pm T_{\kappa}(\psi)}{1 \mp \kappa T_{\kappa}(\phi) T_{\kappa}(\psi)} .
 \end{gather*}


\section{Appendix:  The Hermitian metric}

The hermitian metric
\[ ds^2 = \frac{dw d\overline{w}}{\left( 1 + \ka \left| w \right|^2 \right)^2} \]
 was used in Section~3 to construct conformal models for the Cayley--Klein geometries.

Following Cayley and Klein we can construct a homomorphism from $SL(2,\C)$
to the group of M\"{o}bius transformations as follows.
Let $u$ and $v$ be complex numbers, where the two component vector $\left(
   \begin{matrix}
   u \\
   v
   \end{matrix}
\right) $ will be called a spinor.  If $\left(
   \begin{matrix}
   a & b \\
   c & d
   \end{matrix}
\right)$ is an element of $SL(2,\C)$, then writing
\[
\left(
   \begin{matrix}
   a & b \\
   c & d
   \end{matrix}
\right)
\left(
   \begin{matrix}
   u \\
   v
   \end{matrix}
\right) =
\left(
   \begin{matrix}
   u^{\prime} \\
   v^{\prime}
   \end{matrix}
\right)
\]
we can def\/ine
\[
 w \equiv \frac{u}{v}, \qquad w^{\prime} \equiv \frac{u^{\prime}}{v^{\prime}}
\]
so that
\[ w^{\prime} = \frac{au + bv}{cu + dv} = \frac{aw + b}{cw + d} .
\]
The isometry group of $\varsigma$ with metric $g_1$ is that
given by those transformations belonging to ${\bf Spin}(3)$.  After some tedious algebra we have that
\[
\frac{dw^{\prime} d\overline{w^{\prime}}}{\left( 1 + \ka \left| w^{\prime} \right|^2 \right)^2} =
\frac{dw d\overline{w}}{\left( 1 + \ka \left| w \right|^2 \right)^2}
\]
when
\[
\left(
   \begin{matrix}
   a & b \\
   c & d
   \end{matrix}
\right)
\in SU(2)
\]
so that
\[
ds^2 =
\frac{dw d\overline{w}}{\left( 1 + \ka \left| w \right|^2 \right)^2}
\]
gives the main metric $g_1$ on $\varsigma$.  We have then proved the following lemma.

\begin{lemma}
Those M\"{o}bius transformations that correspond to ${\bf Spin}(3)$
form the isometry group of $\varsigma$ with main metric
\[
g_1 = \frac{dw d\overline{w}}{\left( 1 + \ka \left| w \right|^2 \right)^2} .
\]
\end{lemma}

We would also like to show, following the proof that is given in \cite{BEG99} for the hyperbolic plane, that
\[
d (w_1, w_2) = \Ta^{-1} \left( \left|  \frac{w_2 - w_1}{\ka \overline{w_1}w_2 + 1}  \right| \right)
\]
where $d(w_1, w_2)$ is the Cayley--Klein distance between two points $w_1$ and $w_2$ lying in $\varsigma$.  Let
\[
M(w) = \frac{\alpha w + \beta}{-\ka \overline{\beta} w + \overline{\alpha}}
\]
be a M\"{o}bius transformation where
\[
\left(
   \begin{matrix}
   \alpha & \beta \\
   -\ka \overline{\beta} & \overline{\alpha}
   \end{matrix}
\right) \in SU(2).
\]
without loss of generality $\ka > 0$, so that if $\alpha$, $\beta$,
and $c$ are small positive numbers, then the transformation
\[
[0, c] \longrightarrow \left[ \frac{\beta}{\alpha}, \frac{\beta + \alpha c}{\alpha - \ka \beta c} \right]
\]
induced by $M$ is bijective, and the intersection of the real axis
with $\varsigma$ is a geodesic\footnote{Geodesics of $\varsigma$
are projections of the intersections of planes through the origin
with the unit sphere $\Sigma$: in this case the plane is the $zt$-coordinate plane.}.
Since $M$ is an isometry of $\varsigma$ and distances are additive along a geodesic,
\[
d \left( 0,  \frac{\beta + \alpha c}{\alpha - \ka \beta c} \right) =
d \left( 0, \frac{\beta}{\alpha} \right) +
d \left( \frac{\beta}{\alpha}, \frac{\beta + \alpha c}{\alpha - \ka \beta c} \right) =
d \left( 0, \frac{\beta}{\alpha} \right) + d(0, c) .
\]
Let us def\/ine the quantities
\[
\epsilon = d\left(0,\dfrac{\beta}{\alpha}\right) \qquad \mbox{and} \qquad t = d(0,c)
\]
so that
\[
d \left( 0,  \frac{\beta + \alpha c}{\alpha - \ka \beta c} \right) = \epsilon + t.
\]
Let $g$ denote the inverse of $d: [0,c] \rightarrow [0,t]$, where $d(w)$ is shorthand for $d(0,w)$.
Then\footnote{We can see from the equation below that $g(t) = T_{\ka}(t)$.}
\[
g(t + \epsilon) =
\frac{\beta + \alpha c}{\alpha - \ka \beta c} =
\frac{\frac{\beta}{\alpha} + c}{1 - \frac{\ka \beta}{\alpha}c} =
\frac{g(\epsilon) + g(t)}{1 - \ka g(\epsilon)g(t)}
\]
and so
\[
g(t + \epsilon) -\ka g(t + \epsilon) g(t) g(\epsilon) = g(\epsilon) + g(t)
\]
and then we can divide by $\epsilon$
\[
\frac{g(t + \epsilon) - g(t)}{\epsilon} =
\frac{g(\epsilon)}{\epsilon} \left[ 1 + \ka g(t) g(t + \epsilon)  \right]
\]
and take the limit
\[
\lim_{\epsilon \rightarrow 0^+} \frac{g(\epsilon)}{\epsilon} =
\lim_{\epsilon \rightarrow 0^+} \frac{\frac{\beta}{\alpha}}{\Ta^{-1}\left( \frac{\beta}{\alpha} \right)} =
\lim_{\phi \rightarrow 0^+} \frac{\phi}{\Ta^{-1} (\phi)} =
\lim_{\phi \rightarrow 0^+} \frac{1}{\frac{1}{1 + \ka \phi^2}} =
1.
\]
So $g^{\prime}(t) = 1 + \ka g^2(t)$.  By the inverse function rule for dif\/ferentiation,
\[
d^{\prime}(w) = \frac{1}{1 + \ka w^2}
\]
and so $d(w) = \Ta^{-1}(w)$ as $d(0) = 0$.

If $M$ is the M\"{o}bius transformation given by
\[
M(w) = \frac{c w - c w_1}{c \ka \overline{w_1}w + c}
\]
and where
\[
c = \frac{1}{\sqrt{1 + \ka |w|^2}},
\]
then
\[
M(w_2) \rightarrow \frac{w_2 - w_1}{\ka \overline{w_1} w_2 + 1}
\]
as $w_1 \rightarrow 0$.  Since
\[
d \left( 0, \frac{w_2 - w_1}{\ka \overline{w_1} w_2 + 1} \right) =
d \left( 0, \left| \frac{w_2 - w_1}{\ka \overline{w_1} w_2 + 1} \right| \right)
\]
as rotations are isometries, then $d(w_1, w_2) =$
\[
d \left( 0, \frac{w_2 - w_1}{\ka \overline{w_1} w_2 + 1} \right) =
\Ta^{-1} \left( \frac{w_2 - w_1}{\ka \overline{w_1} w_2 + 1} \right) =
\Ta^{-1} \left( \left| \frac{w_2 - w_1}{\ka \overline{w_1} w_2 + 1} \right| \right) .
\]
So we have proven the following lemma.
\begin{lemma}
If $w_1$ and $w_2$ are two points of $\varsigma$ given the metric $g_1$, then the distance between them is given by
\[
d\left( w_1, w_2 \right) = \Ta^{-1} \left( \left| \frac{w_2 - w_1}{\ka \overline{w_1} w_2 + 1} \right| \right) .
\]
\end{lemma}


\section{Appendix:  Tables}

Tables 13 and 14 are referred to at the end of Section 3, and Tables 15 and 16 are referred to at the end of Section 6.

\subsection*{Acknowledgements}

I wish to thank the referees for their careful reading
of this paper and their suggestions for valuable improvements.

\begin{table}[t]
\centering
\caption{Elements of $SL(2, \C)$ corresponding to $e^{\alpha H}$.}
\vspace{1mm}

\begin{tabular}{l |  l}
\hline
                                           & \tsep{0.5ex}
                                           Elements of $SL(2, \C)$ corresponding to $e^{\alpha H}$ \\  \hline
                                            $\frac{\ka}{\kb}$ is positive &
\tsep{4ex} $\pm \left(
\begin{matrix}
\Cb \left( \sqrt{\frac{\ka}{\kb}} \frac{\alpha}{2} \right) &
\sqrt{\frac{\kb}{\ka}} \Sb \left( \sqrt{\frac{\ka}{\kb}} \frac{\alpha}{2} \right) \\
-\kb \sqrt{\frac{\ka}{\kb}} \Sb \left( \sqrt{\frac{\ka}{\kb}} \frac{\alpha}{2} \right) &
\Cb \left( \sqrt{\frac{\ka}{\kb}} \frac{\alpha}{2} \right)
\end{matrix}
\right)$ \bsep{5ex}\\
$\frac{\ka}{\kb}$ is negative &
$\pm \left(
\begin{matrix}
C_{-\kb} \left( \sqrt{-\frac{\ka}{\kb}} \frac{\alpha}{2} \right) &
\sqrt{-\frac{\kb}{\ka}} S_{-\kb} \left( \sqrt{-\frac{\ka}{\kb}} \frac{\alpha}{2} \right) \\
\kb \sqrt{-\frac{\ka}{\kb}} S_{-\kb} \left( \sqrt{-\frac{\ka}{\kb}} \frac{\alpha}{2} \right) &
C_{-\kb} \left( \sqrt{-\frac{\ka}{\kb}} \frac{\alpha}{2} \right)
\end{matrix}
\right)$ \bsep{5ex}\\
$\ka = 0$                              &
$\pm \left(
\begin{matrix}
1 &  \frac{\alpha}{2} \\
0 & 1
\end{matrix}
\right)$\bsep{3ex}\\
$\ka \neq 0, \kb = 0$            &
$\pm \left(
\begin{matrix}
\Ca \left( \frac{\alpha}{2} \right)  &
\Sa \left( \frac{\alpha}{2} \right) \\
-\ka \Sa \left( \frac{\alpha}{2} \right)  &
\Ca \left( \frac{\alpha}{2} \right)
\end{matrix}
\right)$\bsep{3ex}\\
                                              \hline \hline
\multicolumn{2}{r} \,
Derivatives at $\alpha = 0$ are given by
$\pm \left(\begin{matrix}
0 & \frac{1}{2} \\
-\frac{\ka }{2} & 0
\end{matrix}
\right)$\tsep{2ex}\bsep{2ex}\\
 \hline
\end{tabular}
\end{table}

\newpage

\begin{table}[t]
\centering
\caption{Elements of $SL(2, \C)$ corresponding to $e^{\beta P}$.}
\vspace{1mm}
\begin{tabular}{l |  l}
\hline
\tsep{0.5ex}
                                           & Elements of $SL(2, \C)$ corresponding to $e^{\beta P}$ \\ \hline
$\ka > 0$                            &
\tsep{4ex} $\pm \left(
\begin{matrix}
\Cb \left( \sqrt{\ka} \frac{\beta}{2} \right) &
\frac{i}{\sqrt{\ka}} \Sb \left( \sqrt{\ka} \frac{\beta}{2} \right) \\
 i \sqrt{\ka} \Sb \left( \sqrt{\ka} \frac{\beta}{2} \right) &
\Cb \left( \sqrt{\ka} \frac{\beta}{2} \right)
\end{matrix}
\right)$ \bsep{4ex}\\
$\ka < 0$                            &
$\pm \left(
\begin{matrix}
C_{-\kb} \left( \sqrt{-\ka} \frac{\beta}{2} \right) &
\frac{i}{\sqrt{-\ka}} S_{-\kb} \left( \sqrt{-\ka} \frac{\beta}{2} \right) \\
- i \sqrt{-\ka} S_{-\kb} \left( \sqrt{-\ka} \frac{\beta}{2} \right) &
C_{-\kb} \left( \sqrt{-\ka} \frac{\beta}{2} \right)
\end{matrix}
\right)$ \bsep{4ex}\\
$\ka = 0$                              &
$\pm \left(
\begin{matrix}
1 &  i \frac{\beta}{2} \\
0 & 1
\end{matrix}
\right)$\\
                                            \hline \hline
\multicolumn{2}{r} Derivatives at $\beta = 0$ are given by
$\pm \left(
\begin{matrix}
0 & \frac{i}{2} \\
\frac{\ka i}{2} & 0
\end{matrix}
\right)$\tsep{2ex} \bsep{2ex} \\ \hline
\end{tabular}
\end{table}

{\samepage

\begin{table}[t]
\centering

\caption{The additional basis elements for $sl(2,\C)$ and their one-parameter subgroups in $SL(2,\C)$.}
\vspace{1mm}

 \begin{tabular}{ l || l } \hline
Additional basis  & Corresponding one-parameter subgroup                           \\
elements for $sl(2,\C)$    & in $SL(2,\C)$                                       \\ \hline \hline
\tsep{2ex}
$G_1  =
   \left(
   \begin{matrix}
   0 & 0 \\
   1 & 0
   \end{matrix}
   \right)$ &
  $ \left(
   \begin{matrix}
   1 & 0 \\
   t & 1
   \end{matrix}
   \right)$
 \bsep{2ex}   \\
$G_2 =
   \left(
   \begin{matrix}
   0 & 0 \\
   i & 0
   \end{matrix}
   \right)$ &
  $ \left(
   \begin{matrix}
   1 & 0 \\
   ti & 1
   \end{matrix}
   \right)$
\bsep{2ex}    \\
$D  =   \frac{1}{2}
   \left(
   \begin{matrix}
   1 & 0 \\
   0 & -1
   \end{matrix}
   \right)$ &
  $ \left(
   \begin{matrix}
   e^{\frac{t}{2}} & 0 \\
   0 & e^{-\frac{t}{2}}
   \end{matrix}
   \right)$  \bsep{2.5ex}
 \\ \hline
 \end{tabular}
 \end{table}

 \begin{table}[th]
 \centering
   \caption{The additional basis elements for $sl(2,\C)$ and their corresponding M\"{o}bius transformations.}
   \vspace{1mm}

 \begin{tabular}{ l || l } \hline
Additional basis & Corresponding M\"{o}bius  transformation of $\C$                           \\
elements for $sl(2,\C)$    &                                      \\ \hline \hline
\tsep{2ex}
$G_1  =
   \left(
   \begin{matrix}
   0 & 0 \\
   1 & 0
   \end{matrix}
   \right)$ &
  $ w \mapsto
   \frac{w}{tw + 1} = \frac{1}{t + \frac{1}{w}}
   $
\bsep{2ex}     \\
$G_2 =
   \left(
   \begin{matrix}
   0 & 0 \\
   i & 0
   \end{matrix}
   \right)$ &
  $ w \mapsto \frac{w}{tiw + 1} = \frac{1}{ti + \frac{1}{w}}$
\bsep{2ex}    \\
$D  =   \frac{1}{2}
   \left(
   \begin{matrix}
   1 & 0 \\
   0 & -1
   \end{matrix}
   \right)$ &
  $ w \mapsto e^{t}w $
 \bsep{2ex}\\ \hline
 \end{tabular}
 \end{table}}

\pdfbookmark[1]{References}{ref}
\LastPageEnding

\end{document}